\documentclass[iop, revtex4]{emulateapj}
\usepackage{pdflscape}
\usepackage{amsmath}
\usepackage{epsf}
\usepackage{psfig}
\usepackage [english]{babel}
\usepackage [autostyle, english = american]{csquotes}
\usepackage{chngcntr}
\usepackage{hyperref}
\MakeOuterQuote{"}
\pdfoutput=1 

\usepackage{graphicx}
\usepackage{verbatim}
\usepackage{color}
\usepackage{longtable}
\newcommand{\myemail}{bajoshi@asu.edu}

\newcommand{\pearszrange}{$0.600 \leq {\rm z} \leq 1.235$}
\newcommand{\totalcatalog}{497}

\shorttitle{}
\shortauthors{Joshi et al.}

\received{March 20, 2019}
\accepted{August 6, 2019}

\begin{document}

\title{Spectrophotometric redshifts for $\mathrm{\MakeLowercase{z}\sim1}$ galaxies and predictions for number densities with WFIRST and Euclid}

\author{Bhavin A.\ Joshi\altaffilmark{1}, Seth Cohen\altaffilmark{1}, Rogier A.\ Windhorst\altaffilmark{1}, Rolf Jansen\altaffilmark{1}, Norbert Pirzkal\altaffilmark{2}, Nimish P.\ Hathi\altaffilmark{2}}

\altaffiltext{1}{School of Earth and Space Exploration, Arizona State University, PO Box 871404, Tempe, AZ 85287, USA}
\altaffiltext{2}{Space Telescope Science Institute, 3700 San Martin Drive, Baltimore, MD 21218, USA}
\email{\myemail}

\begin{abstract}
We investigate the accuracy of 4000\AA/Balmer-break based redshifts by combining Hubble Space Telescope ({\it HST}) grism data with photometry. The grism spectra are from the Probing Evolution And Reionization Spectroscopically (PEARS) survey with {\it HST} using the G800L grism on the Advanced Camera for Surveys (ACS). The photometric data come from a compilation by the 3D-HST collaboration of imaging from multiple surveys (notably CANDELS and 3D-HST). 
We show evidence that spectrophotometric redshifts (SPZs) typically improve the accuracy of photometric redshifts by $\sim$17--60\%.  Our SPZ method is a template fitting based routine which accounts for correlated data between neighboring points within grism spectra via the covariance matrix formalism and also accounts for galaxy morphology along the dispersion direction.
We show that the robustness of the SPZ is directly related to the fidelity of the D4000 measurement.
We also estimate the accuracy of continuum-based redshifts, i.e., for galaxies that do not contain strong emission lines, based on the grism data alone ($\sigma^{\rm NMAD}_{\Delta z/(1+z)}{\lesssim}0.06$).
Given that future space-based observatories like WFIRST and Euclid will spend a significant fraction of time on slitless spectroscopic observations, we estimate number densities for objects with $\left| \mathrm{\Delta z/(1+z_{s})} \right| \leq 0.02$. We predict $\sim$700--4400 galaxies/degree$^2$ for galaxies with D4000$>$1.1 and $\left| \mathrm{\Delta z/(1+z_{s})} \right| \leq 0.02$  to a limiting depth of $i_{AB}$=24 mag. This is \emph{especially} important in the absence of an accompanying rich photometric dataset like the existing one for the CANDELS fields, where redshift accuracy from future surveys will rely only on the presence of a feature like the 4000\AA/Balmer breaks or the presence of emission lines within the grism spectra. 
\end{abstract}


\section{Introduction}
Galaxy evolution studies and cosmological measurements require redshift accuracy at the few-percent level or better. In particular, cosmological measurements such as measurements of the baryon acoustic scale \citep[e.g.,][]{Eisenstein2005, Weinberg2013}, and weak lensing tomography \citep[e.g.,][]{Hildebrandt2012}, require redshift accuracy at the percent or better level, and outlier fractions at the sub-percent level \citep{Weinberg2013}. Measurements of galaxy overdensities also require redshifts accurate at the level of a few percent. For example, \citet{Pharo2018}, identify overdensities using redshifts estimated from low-resolution grism spectra combined with broad-band photometry. Accurate redshifts (typically with accuracy of ${\rm \Delta z/(1+z)}{\simeq}0.001$) can be obtained by using high-resolution spectroscopic data that allow for the precise fitting of high-resolution synthetic spectra of stellar populations. These can distinguish between synthetic stellar population models from different regions of parameter space, and simultaneously provide accurate redshifts and stellar population parameters. In practice, it is extremely difficult and very expensive to conduct a large scale spectroscopic survey of faint and distant galaxies that is both unbiased and sufficiently deep to analyze the stellar continua, in order to secure both accurate redshifts \emph{and} detailed stellar population properties. Wide-field large scale ground-based spectroscopic campaigns, e.g., SDSS \citep{York2000}, 6dF \citep{Jones2004}, GAMA \citep{Driver2011}, have been conducted to obtain high-resolution spectra and accurate redshifts. These spectroscopic ground-based surveys, however, are limited to the relatively brighter sources ($\sim 21\text{--}22\ \mathrm{mag}$) at lower redshifts (z$\lesssim$0.5).

In addition, large scale Hubble Space Telescope ({\it HST}) photometric and grism spectroscopic surveys have obtained photometric redshifts accurate to within a few percent, and have led to a better understanding of the stellar populations of a substantial number of galaxies at intermediate and high redshifts (z$>$1.5). Notable examples include the Wide Field Camera 3 (WFC3) Early Release Science (ERS) field \citep{Windhorst2011} and the Cosmic Assembly Near-infrared Deep Extragalactic Survey \citep[CANDELS;][]{Grogin2011, Koekemoer2011} which used imaging from WFC3/IR and the Advanced Camera for Surveys (ACS) on the {\it HST}, while the GRAPES \citep{Pirzkal2004, Pasquali2006, Ryan2007, Hathi2009}, PEARS \citep{Ferreras2009}, and FIGS \citep{Pirzkal2017} surveys invested 40, 200, and 160 {\it HST} orbits, covering 11.6, 119, 18.6 arcmin$^2$, respectively to do slitless spectroscopy with the ACS/G800L and WFC3/G102 grisms. Similarly the 3D-HST survey \citep{Brammer2012, Skelton2014, Bezanson2016}, which is a {\it HST} survey with the WFC3/G141 grism, also invested 248 orbits to conduct WFC3/G141 spectroscopy of the CANDELS fields covering 600 arcmin$^2$.

The 4000\AA\ break is the strongest absorption feature in rest-frame visible spectra of galaxies, particularly in early-type galaxies whose optical light is dominated by older stars, and also to a lesser extent and more varying extent in late-type star forming galaxies \citep{Hathi2009}. The break is expected to be stronger for later stellar spectral types and higher metallicities \citep[see, e.g.,][]{Bruzual1983, Hamilton1985}, and therefore for galaxies dominated by old and metal-rich stellar populations. It is caused by the superposition of multiple absorption features within a narrow wavelength range close to 4000\AA. The H and K absorption lines of \ion{Ca}{2}, at 3969\AA\ and 3934\AA\ respectively, make up a significant part of the amplitude of the 4000\AA\ break, which is why it is sometimes also referred to as the Ca H and K break. The strength of the 4000\AA\ break is an excellent proxy for the age of the stellar population, and for the lack of recent star formation activity \citep[e.g.,][]{Bruzual1983, Hamilton1985, Poggianti1997, Kauffmann2003a, Kauffmann2003b, Hernan-Caballero2013}. Similar to the 4000\AA\ break, the Balmer break at 3646\AA --- caused by strong Balmer absorption lines in younger stellar populations ($\lesssim$0.3 Gyr) --- is also a useful feature for redshift determination. The 4000\AA/Balmer breaks are useful as photometric redshift indicators to achieve accuracy of better than a few percent, particularly if the photometry bands straddle the 4000\AA/Balmer breaks.

In this paper, we combine grism spectra, containing a 4000\AA/Balmer break, and photometric data to derive spectrophotometric redshifts (SPZs) and compare the accuracy of SPZs to redshifts derived from only photometric data (photo-z) and examine the dependence of redshift accuracy on D4000. 

The deep grism data we use in this paper come from the Probing Evolution And Reionization Spectroscopically (PEARS) survey done with the ACS/G800L grism on {\it HST} \citep{Ferreras2009, Straughn2009, Xia2011, Pirzkal2013}. These slitless grism spectra are much lower in resolution (R$\sim$100 for ACS Wide Field Channel WFC/G800L; \citealt{Pasquali2006}) than traditional slit-spectroscopy (which typically have R$\gtrsim$10$^3$), but clearly have higher spectral resolution than broad-band filters. While traditional slit-spectroscopy is inherently restricted to pre-selected objects in the Field-of-View (FoV), slitless grism spectroscopy is capable of providing spectra for nearly \emph{all} sources in the FoV, after accounting for spectral overlap \citep[e.g.,][]{Ryan2018}. 

We use PEARS grism spectra for galaxies within \pearszrange. This redshift range is selected so that the 4000\AA\ break, if present, falls within the ACS/G800L grism wavelength coverage of 0.6--0.95 \micron. We also restrict our sample further by selecting galaxies which have a discernible 4000\AA\ break, i.e., having D4000$\geq$1.1 (see \S\ref{sec:d4000_measurement} and \ref{sec:sample_selection} for details on our measurements and our sample selection). We note that our results are based on grism spectra with a resolution of R$\simeq$100 and a wavelength coverage of $6000 \leq \lambda [\text{\AA}] \leq 9500$. If either or both of these parameters increases i.e., a higher resolution and/or larger wavelength coverage for grism spectra, we expect the SPZ accuracy to improve over photo-z accuracy in absolute terms and also for lower D4000 values, as will be the case for the planned Wide Field InfraRed Survey Telescope (WFIRST) and the Euclid space-based observatories.

This paper is structured as follows: in \S\ref{sec:observations} we provide details of the slitless spectroscopy data and the PEARS survey. In \S\ref{sec:break_and_sample}, we explain the methods used for measurements of the 4000\AA\ break and our sample selection, and in \S\ref{sec:fitting}, we describe our SED fitting procedure for estimating redshifts. In \S\ref{sec:accuracy}, we evaluate the dependence of our photo-z, grism-z, and SPZ accuracy with D4000 by comparing to ground-based spectroscopic redshifts. In \S\ref{sec:num_density}, we provide the number density predictions based on continuum derived redshifts for observations by future observatories, such as WFIRST and Euclid. We conclude in \S\ref{sec:conclusion}. Wherever needed, we used the following cosmology: a flat Universe with H$_0 = 67.4\; \mathrm{km\,s^{-1}\,Mpc^{-1}}$, $\Omega_m = 1 - \Omega_\Lambda = 0.315$, from the Planck 2018 results \citep{Aghanim2018}. All magnitudes quoted in this paper are AB magnitudes \citep{Oke1983}.

\section{Observations}
\label{sec:observations}

\begin{figure*}[ht!]
\centering
\includegraphics[width=0.5\textwidth]{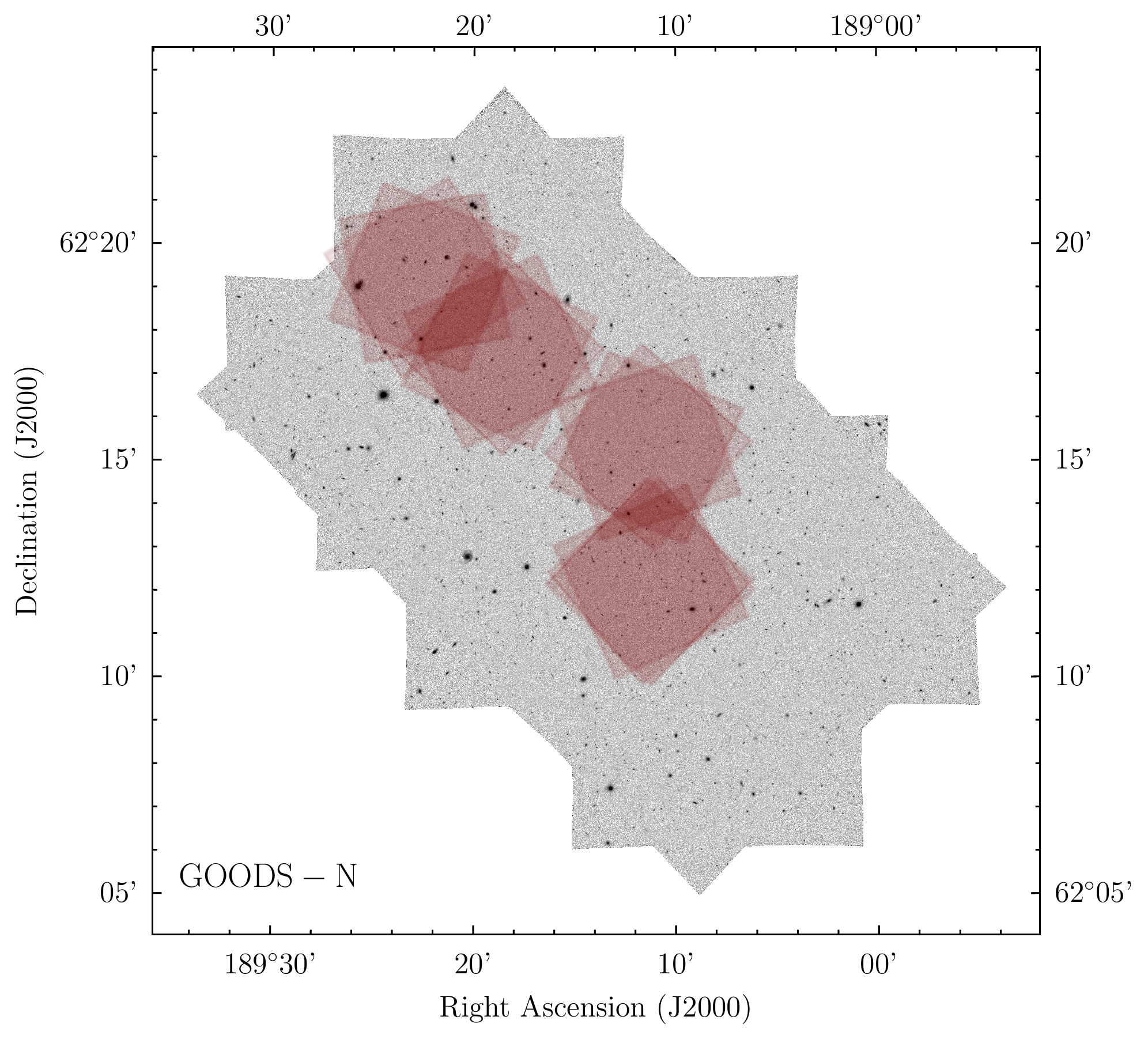}
\includegraphics[width=0.49\textwidth]{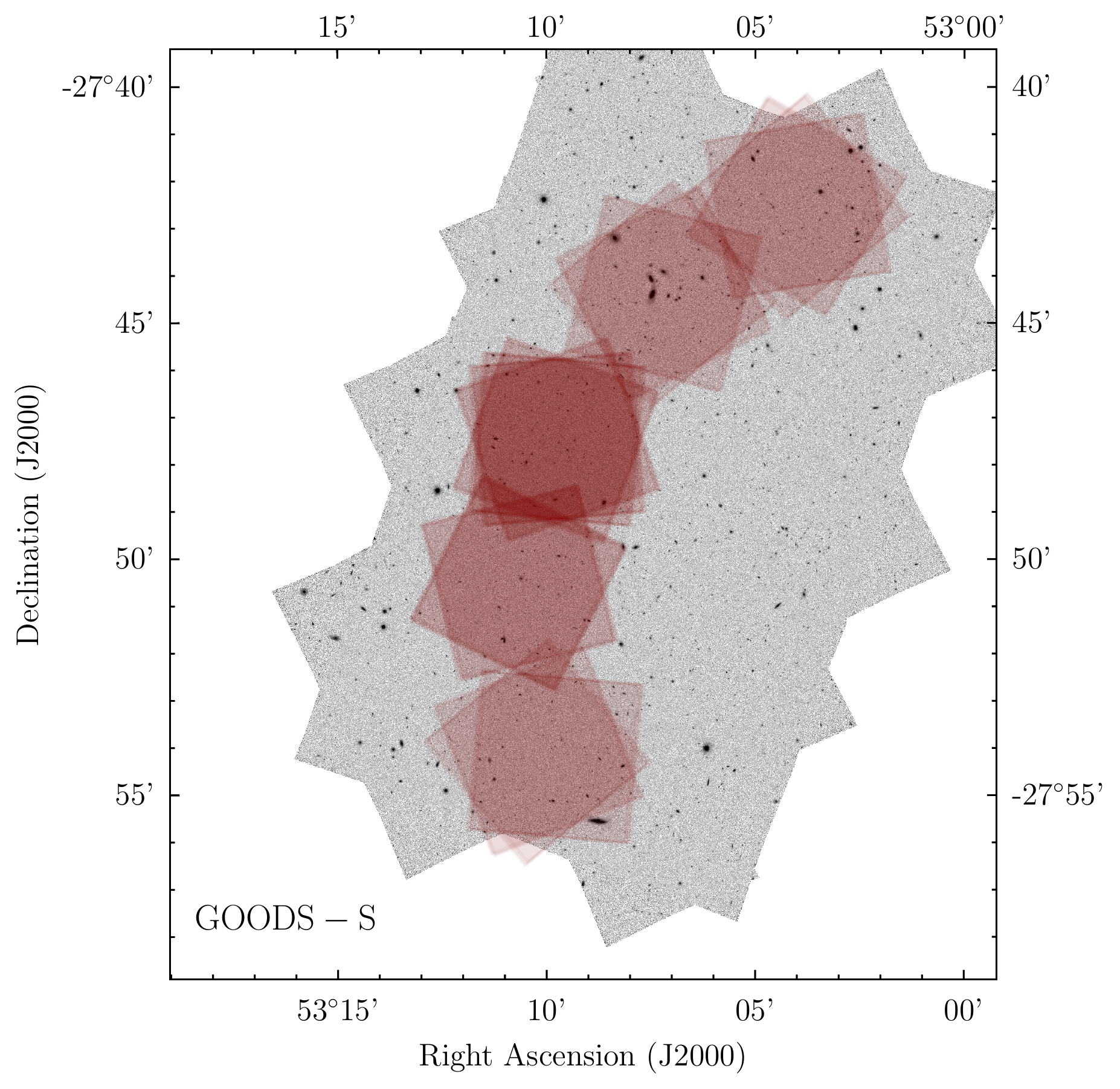}
\caption{The coverage of the PEARS survey within the GOODS fields. The red pointings, superimposed on a HST/ACS/F606W mosaic from the 3D-HST collaboration, show the primary coverage of the PEARS survey with the ACS G800L grism.}
\label{fig:goods_overlay}
\end{figure*}

We use slitless spectroscopy obtained with the {\it HST} as part of the PEARS survey (GO~10530; PI - S. Malhotra). The PEARS survey was awarded 200 orbits in Cycle~14 to cover 8 fields in the Great Observatories Origins Deep Survey (GOODS) North (GOODS-N) and South (GOODS-S) regions \citep{Giavalisco2004a} to a depth of $z'_{\rm AB}$$\le$ 27 mag with the ACS/WFC G800L grism. A ninth ultra-deep field to $z'_{\rm AB}$$\lesssim$ 28 mag overlaps the Hubble Ultra Deep Field \citep[HUDF;][]{Beckwith2006}.
The total area covered is ${\sim}119\, \mathrm{arcmin}^2$. The G800L grism delivers +1$^{st}$ order spectra with a dispersion of about 40\AA\ per pixel \citep{Pasquali2006}. The best resolution of R$\sim$100 is achieved for point sources, but for most of the sources considered in this paper (see \S\ref{sec:sample_selection} for details on our selected sample), which are spatially extended, the effective resolution is lower because the spectrum is convolved with the object morphology along the dispersion direction \citep{Pasquali2001}.

The ACS G800L grism nominally covers 0.55--1.05 \micron\ (for spectra dispersed in the positive first order). The useful wavelength range, where the throughput of the grism exceeds 10\%, is 0.60--0.95 \micron. While somewhat dependent on the exact bandpasses used to measure the break strength, this allows one to trace the 4000\AA\ break uninterrupted from z\,$\simeq$\,0.600 to z\,$\simeq$\,1.235\,.
We refer to \S\ref{sec:d4000_measurement} for the definition of the break indices and for the justification of our use throughout this paper of D4000 over D$_{\rm n}$4000. If the D$_{\rm n}$4000 index is used, instead of D4000, then the redshift range is $0.558 \leq {\rm z} \leq 1.317$.

Fig. \ref{fig:goods_overlay} shows the footprints of the PEARS pointings within the GOODS-N and GOODS-S regions. Each of the 8 deep PEARS pointings were observed for a total of 20 {\it HST} orbits, while the ultra-deep pointing on the HUDF totaled 40 orbits.  In order to mitigate contamination of galaxy spectra that may (partially) overlap at any given dispersion direction, each pointing was visited at three different position angles (PAs), except for 2 fields in GOODS-S which were observed at four PAs.  Direct images through the ACS/WFC F606W filter were taken to astrometrically align the G800L grism exposures and to provide the zeropoint of the grism wavelength solution.

For more details about slitless spectroscopy with the ACS/G800L grism and its data reduction the reader is referred to \citet{Pirzkal2004} and \citet{Pasquali2006}. We refer to \citet{Pirzkal2013} for a description of the reduction and analysis specific to PEARS data. \citet{Pasquali2006} describe the basics of ACS grism observations and the strategy used to calibrate grism observations in orbit.

\section{4000\AA\ Break measurement and Sample Selection}
\label{sec:break_and_sample}
\subsection{D4000 Measurement}
\label{sec:d4000_measurement}
We measure the 4000\AA\ break in the rest-frame of the galaxies that were in the PEARS survey, and which also had measured photometry from the 3D-HST and CANDELS surveys. The galaxies are distributed over the redshift range \pearszrange. In this work, the 4000\AA\ break is measured by the D4000 \citep{Bruzual1983, Hamilton1985}, as opposed to the D$_{\rm n}$4000 index (\citealt{Balogh1999}; see below). The D4000 index \citep[see, e.g.,][]{Bruzual1983, Hamilton1985} measures the ratio of the integrated continuum flux density (in f$_\nu$ units) in the bandpass from 4050\AA\ to 4250\AA\ to the integrated continuum flux density in the bandpass from 3750\AA\ to 3950\AA\ (Eq.\ \ref{eq:d4000}). The D$_{\rm n}$4000 index, where the n stands for ``narrow'', measures the ratio of the integrated continuum flux density in the bandpass from 4000\AA\ to 4100\AA\ to the flux in the bandpass from 3850\AA\ to 3950\AA\ (Eq.\ \ref{eq:dn4000}).

    \begin{equation}
    \label{eq:d4000}
    \mathrm{D}4000 = \int^{4250\text{\AA}}_{4050\text{\AA}} f_\nu d\lambda / \int^{3950\text{\AA}}_{3750\text{\AA}} f_\nu d\lambda
    \end{equation}
    
    \begin{equation}
    \label{eq:dn4000}
    \mathrm{D_n4000} = \int^{4100\text{\AA}}_{4000\text{\AA}} f_\nu d\lambda / \int^{3950\text{\AA}}_{3850\text{\AA}} f_\nu d\lambda
    \end{equation}

The narrower definition, D$_{\rm n}$4000, is less sensitive to reddening effects, and is used relatively more in recent literature \citep[e.g,][]{Li2015, Zahid2017}. Because we are using low-resolution grism spectra, the measurement of D$_{\rm n}$4000 is more likely to be less accurate. This is because the break must fall in a relatively narrower wavelength range (as compared to D4000) for the measurement to be accurate. Since the spectrum must also be deredshifted before measuring the break strength, this ``narrow'' constraint also requires the redshift estimate to be more accurate in order for the D$_{\rm n}$4000 measurement to be robust. Furthermore, there are fewer flux measurements within each wavelength bin to integrate over while using D$_{\rm n}$4000 compared to D4000, which could again lead to the D$_{\rm n}$4000 to be less robust than D4000. Therefore, when comparing the accuracy of spectrophotometric redshifts with photometric redshifts for different 4000\AA\ break strengths, we prefer to measure these break strengths in the grism spectra of our galaxies using the D4000 index rather than the D$_{\rm n}$4000 index. Appendix \ref{app:d4000_err} provides details on the error analysis for the D4000 measurement.

\subsection{Sample Selection}
\label{sec:sample_selection}
The sample of galaxies used in this paper comes from the public master catalog of galaxies and their grism spectra released by the PEARS survey\footnote{\url{https://archive.stsci.edu/prepds/pears/}}. This catalog contains 9551 galaxies (4082 in GOODS-N and 5469 in GOODS-S). We matched the PEARS master catalogs (using a matching radius of 0\farcs3) with photometry catalogs from the 3D-HST survey \citep{Skelton2014} and our catalog of ground-based spectroscopic redshifts in the GOODS regions. This gives us all galaxies which have measured photometry, grism spectra, and ground-based spectroscopic redshifts. The matching with the ground-based spectroscopic redshift catalog is done so that we can check the accuracy of our redshifts (see \S\ref{sec:accuracy}). The 3D-HST photometry catalog contains photometry from multiple ground and space-based surveys. We use 12-band photometry from $u$-band (ground-based) to 8$\mu$m ({\it Spitzer} IRAC). We refer the reader to the 3D-HST photometry paper for details on the imaging sources (Table 3 in \citealt{Skelton2014}). This matching results in a sample of 1863 galaxies.

\begin{deluxetable}{c p{4cm} p{3cm}}
\centering
\tablecaption{Summary of sample selection cuts \label{tab:sample}}
\tablehead{
\colhead{} & \colhead{Selection cut} & \colhead{Number of galaxies} \\
\colhead{} & \colhead{} & \colhead{remaining after cut}
}
1. & PEARS master catalog & \hspace{1cm} 9551 \\
2. & Matching with 3D-HST photometric catalog and ground-based spectroscopic redshift catalog & \hspace{1cm} 1863 \\
3. & Redshift cut i.e., $0.600\leq \mathrm{z_{\rm spec}} \leq 1.235$ & \hspace{1cm} 790 \\
4. & Combined cuts: & \hspace{1cm} 602 \\
 & a) $\mathcal{N}>10$ & \\
 & b) Contamination $<33\%$& \\
5. & Final sample for which we measured photometric, grism, and spectrophotometric redshifts. &  \hspace{1cm} \totalcatalog \\
 & a) 1.1$\leq$D4000$<$2.0 &  \\
 & b) None/incomplete flux measurements within D4000 bandpass & \\
 & c) Remove worst quality spectroscopic redshifts & \\
\enddata
\end{deluxetable}

We then applied a redshift cut of \pearszrange\ to the ground-based spectroscopic redshifts to get galaxies which could contain a 4000\AA\ break in their grism spectra. This results in a sample of 790 galaxies. A note on GOODS-N astrometry is in order here: because the PEARS catalogs were made with pre v2.0 ACS GOODS images, before matching the PEARS catalog with 3D-HST and ground-based spectroscopic redshift catalogs, we also corrected for the known offset in the declination of pre v2.0 ACS images for GOODS-N. This offset is $\sim$0.3 arcsec (see the readme file for v2.0 ACS images\footnote{\url{https://archive.stsci.edu/pub/hlsp/goods/v2/h_goods_v2.0_rdm.html}}).

\begin{figure}
\includegraphics[width=0.47\textwidth]{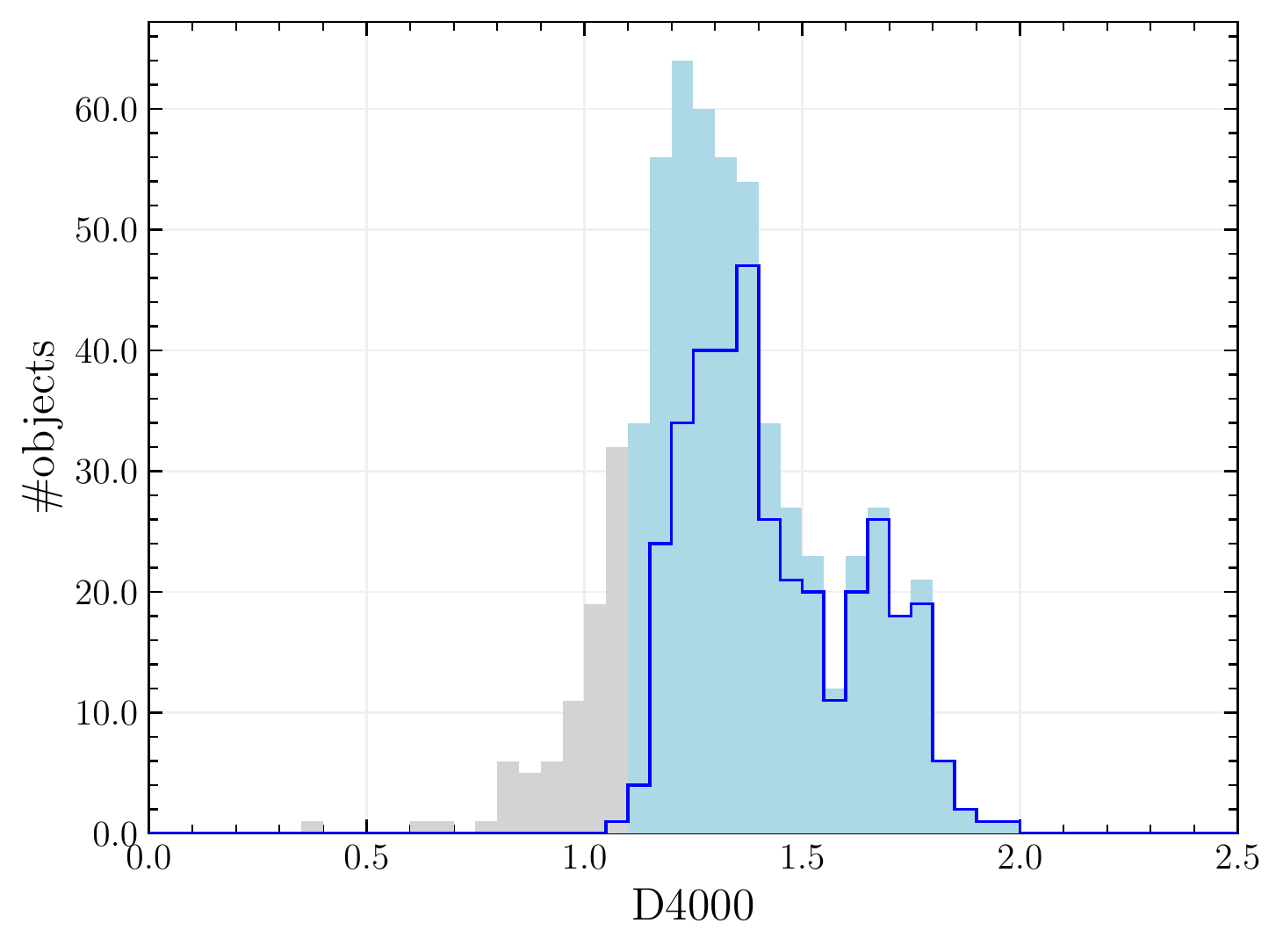}
\caption{The distribution of D4000 for all 602 galaxies within our redshift range and which also passed our Net Spectral Significance and contamination cuts. The light blue shaded area shows the D4000 range for galaxies included in our final sample, i.e., D4000$\geq$1.1. The overlaid blue histogram shows the distribution of those galaxies that have a 3$\sigma$ or better measurement of D4000, i.e., $({\rm D4000} - 1.0)/{\rm \sigma_{D4000}} \geq 3.0$ (see \S\ref{sec:accuracy}).}
\label{fig:d4000_hist}
\end{figure}

We also apply a cut on the Net Spectral Significance, $\mathcal{N} \geq 10$. Briefly, the Net Spectral Significance is a proxy for the useful information content within a spectrum. For example, from \citet{Pirzkal2004}, $\mathcal{N}>8.5(n_{\rm pix}/100)^{1/2}$ corresponds to the detection of at least a 3$\sigma$ signal in the grism data; where $n_{\rm pix}$ is the number of independent spectral elements. In our case, we typically have $n_{\rm pix}$$\sim$88, which implies that a value of $\mathcal{N}>8$ corresponds to at least a 3$\sigma$ detection of signal in the grism data. We refer the reader to appendix \ref{app:netsig} for the definition and also to \citet{Pirzkal2004} for details. We also reject galaxies with excessive contamination (as measured by the PEARS pipeline reduction) -- defined here as any galaxy which has more than 33\%\ of its continuum flux contaminated i.e., likely coming from its line-of-sight neighbors \citep{Pirzkal2004, Pirzkal2013, Pirzkal2017}. We also reject an additional 5 galaxies that have a D4000 error larger than 0.5. These cuts give us 602 galaxies. 
Finally, before we run the code to estimate the three types of redshifts (i.e., photometric, grism, and spectrophotometric), we restrict the range to D4000$\geq$1.1. This ``color'' cut is used to remove grism spectra of galaxies that would not be useful in determining redshifts, since they do not contain a discernible 4000\AA\ break. This brings the final sample of galaxies for which we estimate redshifts to \totalcatalog\ galaxies. Table \ref{tab:sample} summarizes our selection cuts. Figure \ref{fig:d4000_hist} shows the distribution of D4000 in our sample (note that this figure includes galaxies with D4000$<$1.1 to clearly show the distribution).

\section{SED Fitting Procedure}
\label{sec:fitting}
\subsection{Template library}
We use the Bruzual and Charlot (2003; hereafter BC03) library of stellar population synthesis (SPS) models \citep{Bruzual2003} to compare with the observed grism and photometric data of a galaxy to infer its redshift. The synthetic spectra include models with 3 different star formation histories (SFHs): 1) instantaneous burst also referred to as Simple Stellar Populations (SSPs); 2) exponentially declining SFH also referred to as Composite Stellar Populations (CSPs); and 3) constant SFH, where the upper limit of the time scale ($\tau$) on our exponentially declining models is $\sim$63 Gyr. This is much older than the current age of the Universe, so that this model effectively has a constant SFH. 

We generate a grid of templates with the age, metallicity, dust extinction (as measured by $\mathrm{A_V}$) and SFH as parameters. All the models are normalized to form a total stellar mass of 1 M$_\odot$. The models are restricted to an age range of 10 Myr to 7.95 Gyr. This upper limit is decided by the age of the Universe at the lowest value in our redshift range, i.e., this is the oldest possible age for any galaxy in our sample. However, while fitting each individual galaxy, the age of the model is restricted to be less than the age of the Universe, depending on the redshift of the galaxy under consideration. The SSP models have 6 metallicity values: 0.005\,Z$_\odot$, 0.02\,Z$_\odot$, 0.2\,Z$_\odot$, 0.4\,Z$_\odot$, Z$_\odot$, and 2.5\,Z$_\odot$. The CSP models, however, are restricted to solar metallicity values for the sake of computational efficiency. For the exponentially declining SFHs we use a grid for the $e$-folding time $\tau$ (in Gyr) that has a range of $-2<\log{\tau}<+2$ and a step-size $\Delta\log(\tau)$ of 0.02. For a screen of dust, the optical depth, $\tau_{V}$, is related to visual dust extinction, $A_V$, by $A_V = 1.086\tau_{V}$.  For $\tau_{V}$ we adopt a grid with a range of $0.0\leq\tau_{V}\leq2.9$ and a step size of 0.2. The BC03 models use the prescription given by \citet{Charlot2000} to include the effect of dust extinction on the stellar light. The wavelength range for all the models generated by BC03 is 91\AA\ to 160$\mu$m. The total number of templates used is 37761.

Since the BC03 templates do not contain emission lines, we manually add emission lines to the model spectra. Following the prescription given by \citet{Anders2003}, we relate the number of Lyman continuum photons (N$_\mathrm{Lyc}$) and the strength of non-Hydrogen emission lines to the H$\beta$ line strength.
\begin{equation}
    \mathrm{f(H\beta)} = 4.757 \times 10^{-13}\, .\, \mathrm{N_{Lyc}}
\end{equation}
For each template, the BC03 code gives N$_\mathrm{Lyc}$ as one of its output parameters, and this allows us to get the H$\beta$ and metal emission line fluxes. The ratios of the Hydrogen recombination lines are related to the H$\beta$ flux as given by \citet{Hummer1987}, assuming ISM conditions of $n_e=10^2$cm$^{-3}$ and $T_e=10^4$K and Case B recombination. For the sake of computational efficiency we only include typically observed optical emission lines such as H$\alpha$, H$\beta$, H$\gamma$, H$\delta$, [MgII]$\lambda$2800, [OII]$\lambda$3727, [OIII]$\lambda\lambda$4959,5007, [NII]$\lambda\lambda$6548,6583, and [SII]$\lambda\lambda$6716,6731.

\subsection{Fitting and error estimation}
The procedure we follow to arrive at the best fit BC03 model is a $\chi^2$ minimization method which accounts for correlated data in the grism spectrum. For every galaxy, we compare the entire model set to the observed data of the galaxy to get a $\chi^2$ value for each model. The $\chi^2$ statistic is defined by,

\begin{equation}
\label{eq:chi_square}
\chi^2 = \left( F - \alpha M \right)^T \, C^{-1} \, \left( F - \alpha M \right),
\end{equation}

where $F$ and $M$ are the flux and model SED vectors, respectively, which are in flux density units (i.e., erg\,s$^{-1}$\,cm$^{-2}$\,\AA$^{-1}$), and $C^{-1}$ denotes the inverse of the covariance matrix. There is only a single free parameter for each model, the vertical scaling factor $\alpha$, the value of which is found by  finding where the first order derivative of $\chi^2$ vanishes: 

\begin{equation}
\label{eq:alpha}
\frac{\partial \chi^2}{\partial \alpha} = 0
\Rightarrow
\alpha = \frac{\sum_{ij} \left( F_i\,M_j + F_j\,M_i \right) / \sigma^2_{ij}}{2\sum_{ij} M_i\, M_j / \sigma^2_{ij}}
\end{equation}

Here, $1/\sigma^2_{ij}$ is the `ij'th element in the inverse of the covariance matrix. On the diagonal of the covariance matrix this corresponds to the variance on each individual flux point in the observed data. Details on the covariance matrix estimation (to account for correlated data in the extracted slitless spectra) and a brief explanation for Eq.\ \ref{eq:alpha} are given in Appendix \ref{app:covmat}. To arrive at the redshift estimate we choose the redshift corresponding to the best-fit model which has an age that does not exceed the age of the Universe at that redshift.

This procedure remains the same in essence regardless of the redshift that is being computed, i.e., photometric, grism, or spectrophotometric. To estimate photometric redshifts, one needs to (i) redshift the high-resolution model spectra, and (ii) compute fluxes for the redshifted models through all filters for which photometry is available, before the comparison with the photometric data is performed. The only simplification in the case of photometric redshifts is that the covariance matrix is taken to be diagonal for photometric data, since the CANDELS/{\it HST} filters are all essentially non-overlapping and therefore largely independent. In the case of grism redshift estimates, one must (i) redshift the high-resolution model spectra, (ii) convolve the models with the Line Spread Function (LSF), and (iii) resample the redshifted, convolved model spectra to the grism spectral dispersion of 40\AA\ per pixel. Finally, when estimating spectrophotometric redshifts from the combination of grism and photometric data, we apply all these modifications to the models before performing the $\chi^2$ minimization.

We consider a range of possible redshifts for each galaxy within $0.30 \leq {\rm z} < 1.50$ with a step size of $\Delta {\rm z}=0.01$. We experimented with $\Delta {\rm z}=0.005$ and found no significant improvement in the redshift estimates at significant cost of computing time. The larger redshift range is required to properly sample the redshift probability distribution, p(z) curve, in cases where the galaxy redshift is at the edge of the redshift range over which the 4000\AA\ break is visible, i.e., \pearszrange. The redshift in step (i) above comes from iterating over all the redshifts in this range, i.e., for each redshift in this range we carry out the three steps mentioned above, and construct a map of $\chi^2$ values. 

\begin{figure*}
\centering
\includegraphics[width=0.48\textwidth]{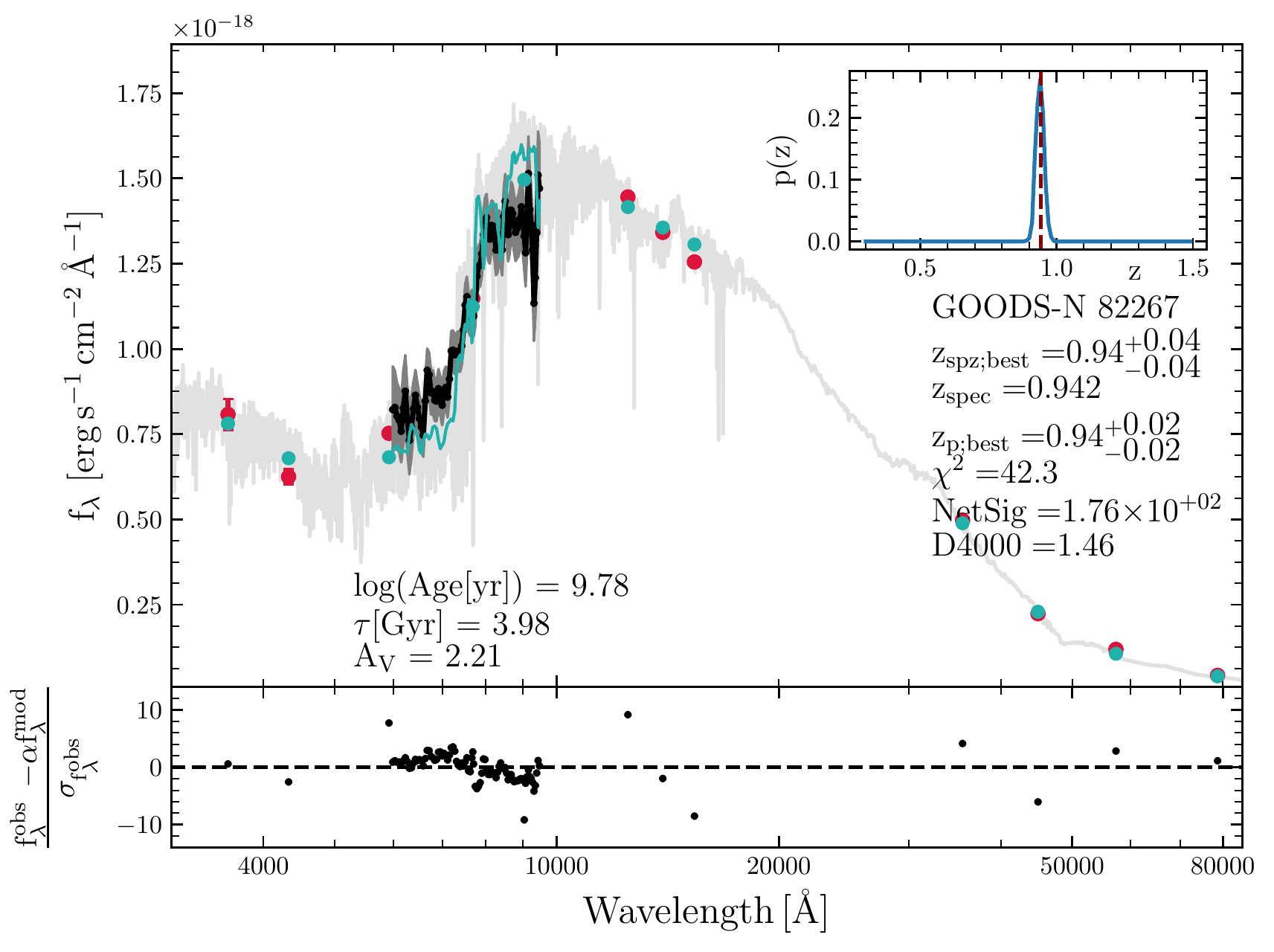}
\includegraphics[width=0.48\textwidth]{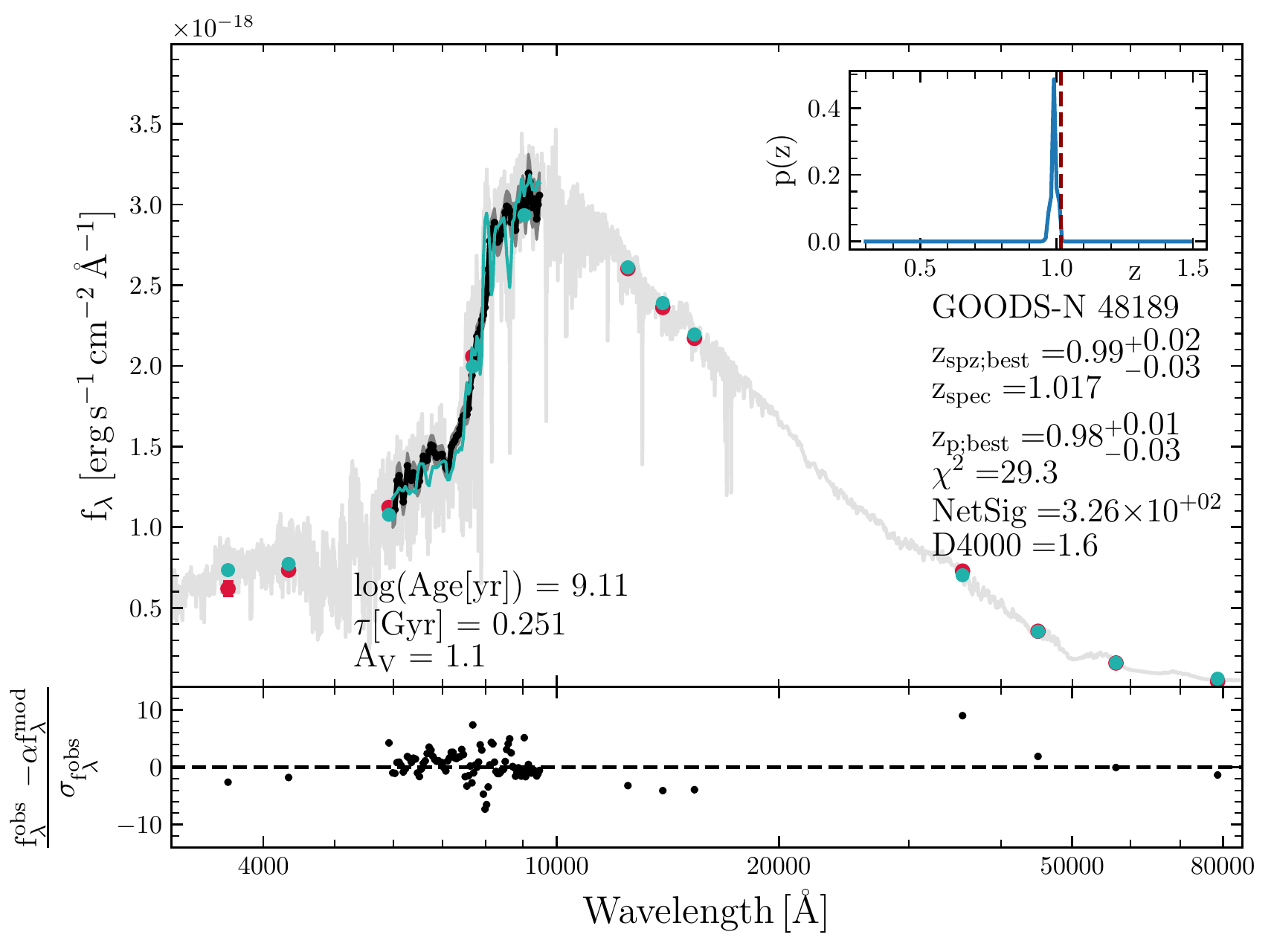}
\includegraphics[width=0.48\textwidth]{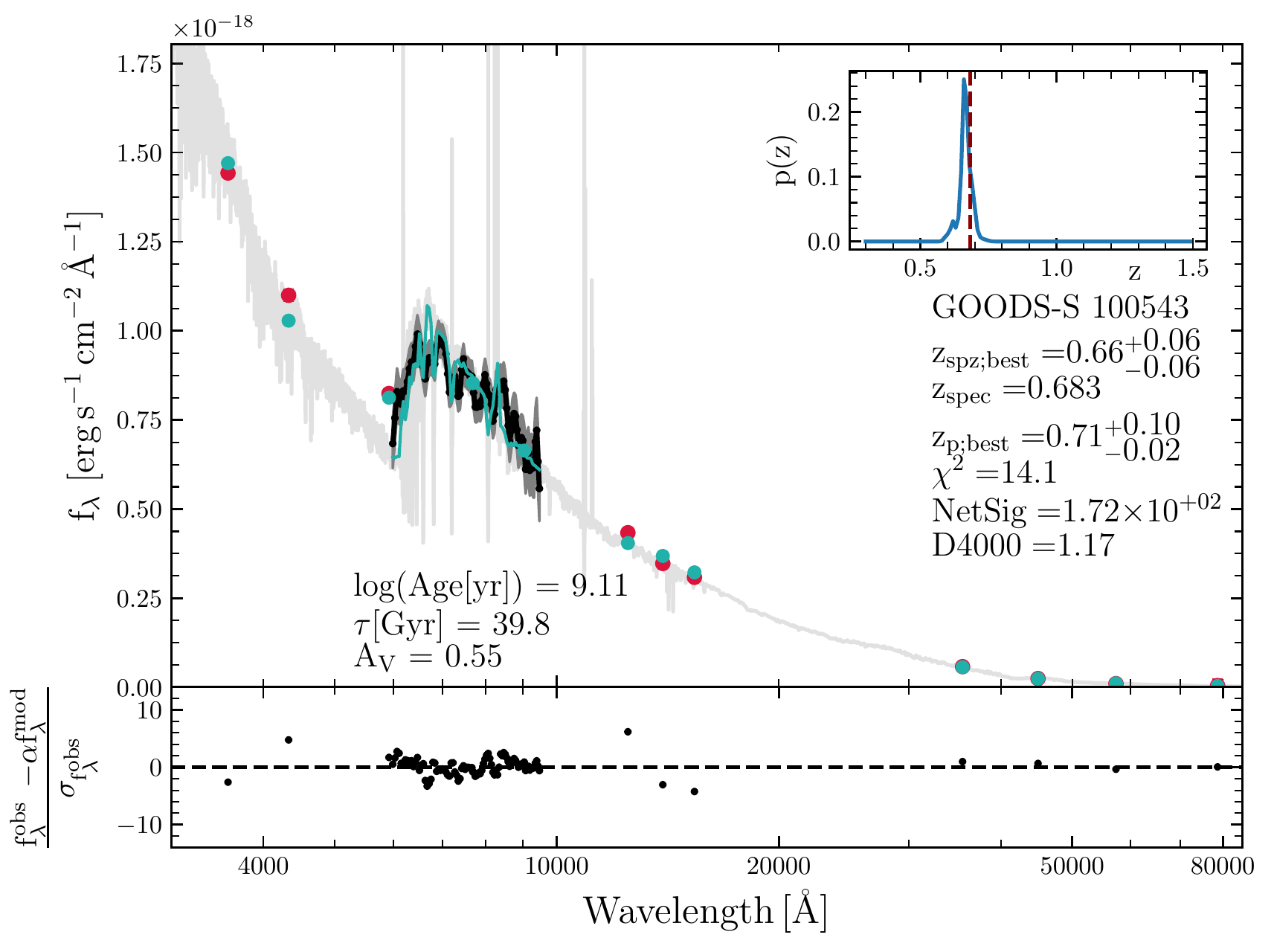}
\includegraphics[width=0.48\textwidth]{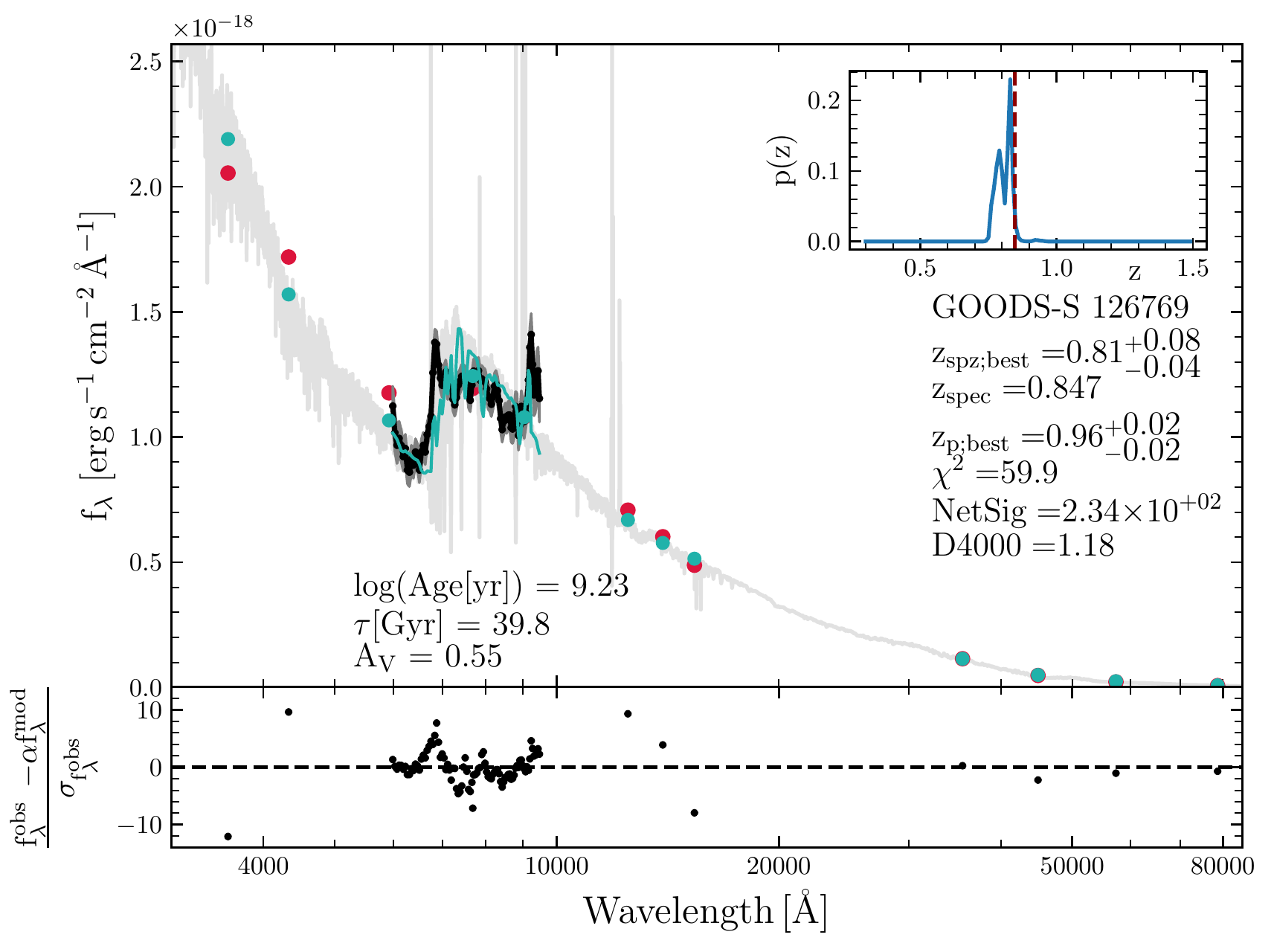}
\caption{Example spectra and fit residuals from our SPZ fitting procedure for four galaxies in our sample at various D4000 values. The black solid line is the grism data with error bars shown in gray. The observed photometry is shown as red points with error-bars. The light gray line is the best fit high-resolution BC03 model at the estimated redshift. The light green solid line and points are the best fit BC03 model downsampled to the grism redshift and the model photometry. The ``best'' spectrophotometric redshift from our code, which is the redshift corresponding to the minimum $\chi^2$, is shown on the plot legend along with the ground-based spectroscopic redshift as well as the photometric redshift. 
The p(z) curve for the galaxy is shown in the upper right corner as an inset figure with the ground-based redshift shown as a red dashed line. 
The plot legend also shows the galaxy ID, Net Spectral Significance ($\mathcal{N}$), D4000 (based on the ground-based spectroscopic redshift), and other derived parameters from the fitting routine.
The bottom panels show the residuals for the fit, i.e., $(f_{\lambda}^{\rm obs}-f_{\lambda}^{\rm model}) / \sigma_{f_{\lambda}^{\rm obs}}$. Note that the flux axis is plotted here in f$_\lambda$ units for visual clarity since the 4000\AA\ break is more prominent in f$_\lambda$ units, but that the D4000 measurement is done in f$_\nu$ units.
}
\label{fig:example}
\end{figure*}

The convolution of the models with the LSF of the galaxy is done to take into account the effects of the morphology of an object on its spectrum. Because of the absence of a spectroscopic slit in grism data, the orientation of the object with respect to the dispersion direction can cause the LSF and hence the resulting spectrum to be quite different at different position angles. This knowledge of the LSF is important because the LSF will cause any absorption or emission features that might be present --- like the 4000\AA\ break --- to be diluted, and therefore cause a significant variation in the measurement of indices such as D4000. This effect is even more pronounced for an index with narrower wavelength bandpasses, such as D$_{\rm n}$4000. In our case, for the PEARS data, we evaluate the Net Spectral Significance, $\mathcal{N}$ (appendix \ref{app:netsig}), for the spectrum at each position angle and select the spectrum with the highest $\mathcal{N}$ to compare to the BC03 SPS models. 

Apart from the LSF convolution, it is also necessary that the SED models and the grism data are sampled to the same spectral resolution before being compared to find a best fitting model. The spectral resolution of the BC03 models is 3\AA\ in the range $3200 < \lambda\,[\text{\AA}] < 9500$, and lower outside this range \citep{Bruzual2003}. A mismatch in resolution would lead to an improper comparison between the high-resolution model and the low-resolution data, because spectral features in the model would remain much sharper than in the data yielding larger values of the $\chi^2$ statistic that we are attempting to minimize. The models are therefore re-sampled to the wavelength resolution of the grism spectrum. This re-sampling of the model is done simply by taking the average flux of all points in the model that fall within the wavelengths of two adjacent points on the grism spectrum. This is done for all points within the grism spectrum wavelength coverage. For the purposes of finding a best-fit model, we only consider the part of the model spectrum that has the same wavelength range as the grism spectra.

\begin{figure*}
\centering
\includegraphics[width=0.9\textwidth]{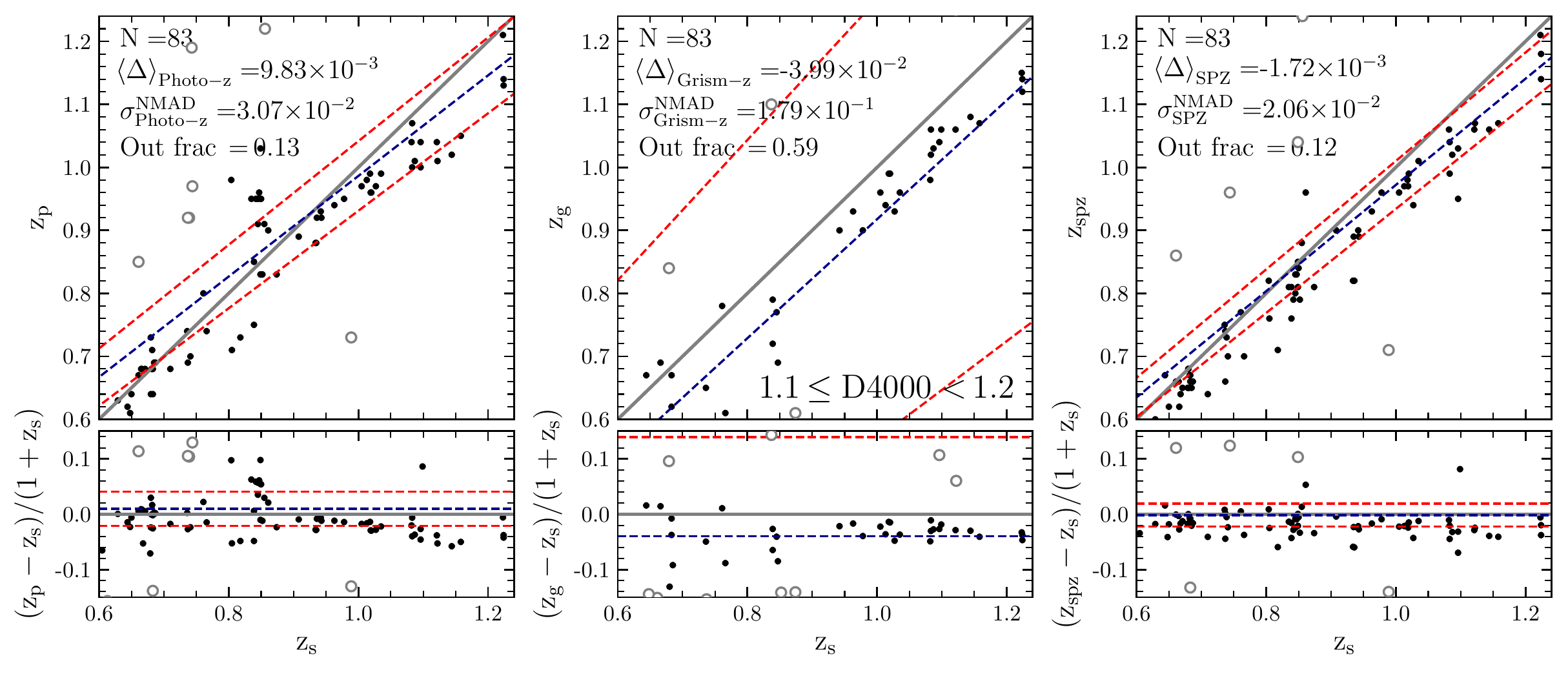}
\includegraphics[width=0.9\textwidth]{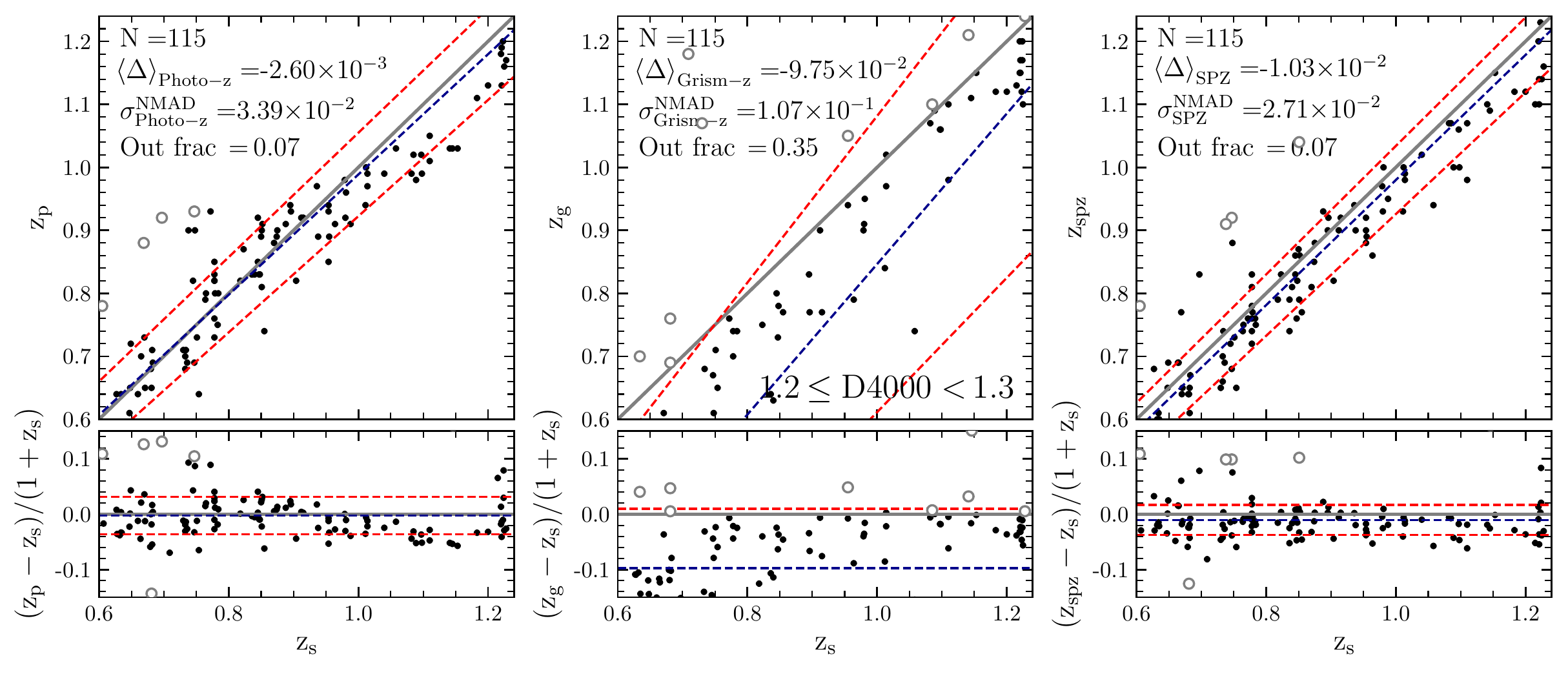}
\includegraphics[width=0.9\textwidth]{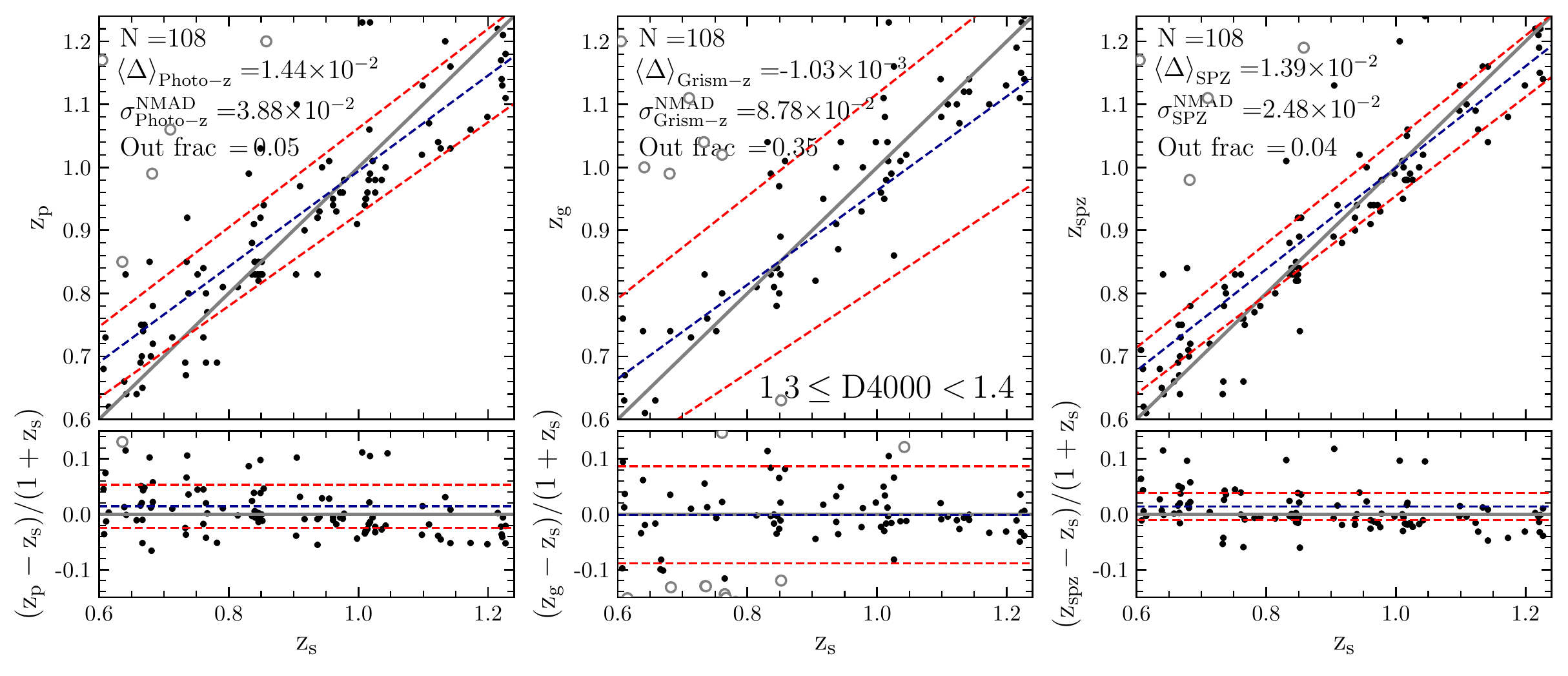}
\caption{Photo-z (z$_{\rm p}$), grism-z (z$_{\rm g}$), and SPZ (z$_{\rm spz}$) redshift accuracy for different D4000 bins by comparing to ground-based spectroscopic redshifts (z$_{\rm s}$). Each row on this figure contains galaxies that fall in the D4000 range shown in the middle subplot. Within each row are three subplots: photo-z on the left, grism-z in the middle, and SPZ on the right. The top panels within each subplot show each of the three redshifts vs.\ the ground-based spectroscopic redshift. The bottom panels show the residuals, i.e., ($\mathrm{z_* - z_{\rm s}}$)/(1+z$_{\rm s}$), where z$_*$ is either z$_{\rm p}$, z$_{\rm g}$, or z$_{\rm spz}$. The D4000 bins sizes are steps of 0.1, except at the two bins with largest D4000 (in Figure \ref{fig:z_comparison_highd4000}), where we have a larger bin size to get a comparable number of galaxies in each bin. The gray solid line in the top panels is the 1:1 line. The blue and red dashed lines are the mean and $\pm$1$\sigma_\mathrm{NMAD}$ spread, respectively. The gray open circles are $>$3$\sigma$ outliers.}
\label{fig:z_comparison_lowd4000}
\end{figure*}

\begin{figure*}
\centering
\includegraphics[width=0.9\textwidth]{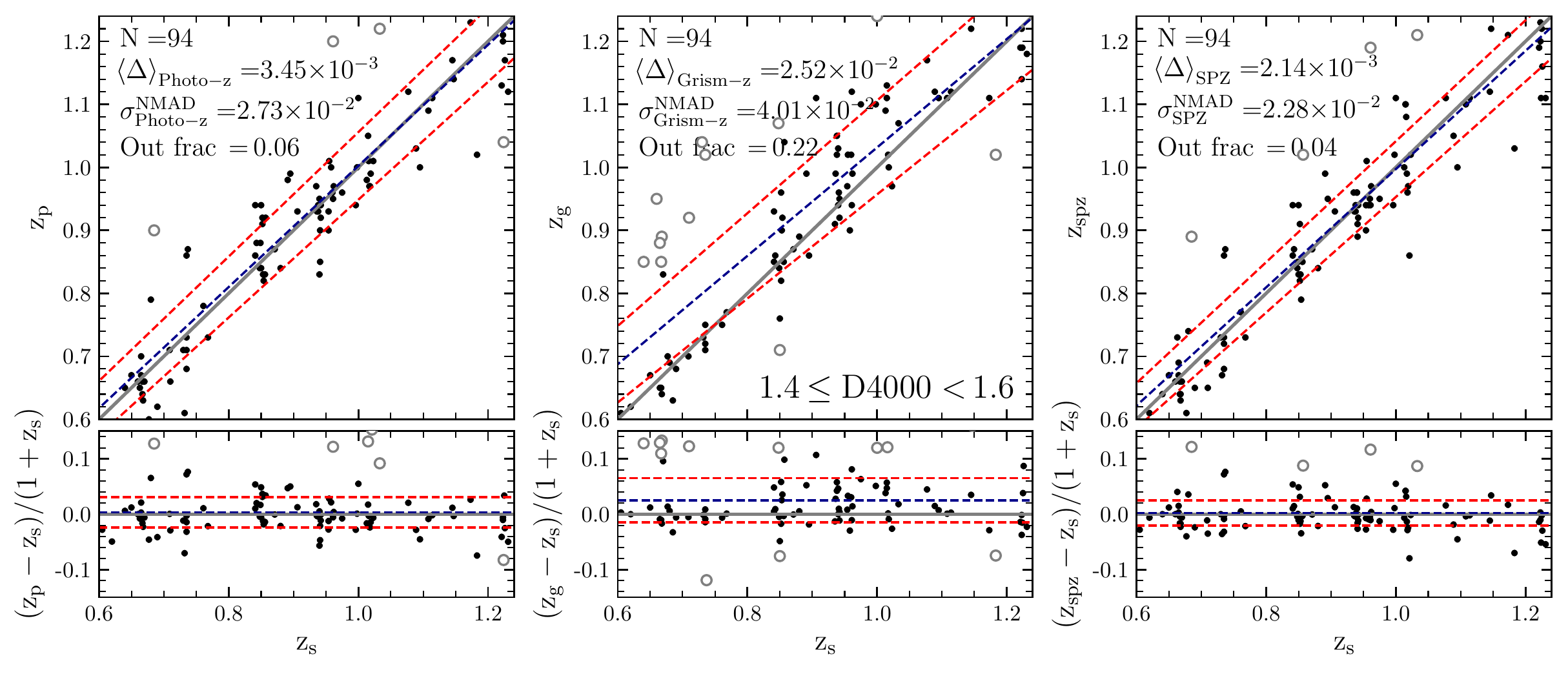}
\includegraphics[width=0.9\textwidth]{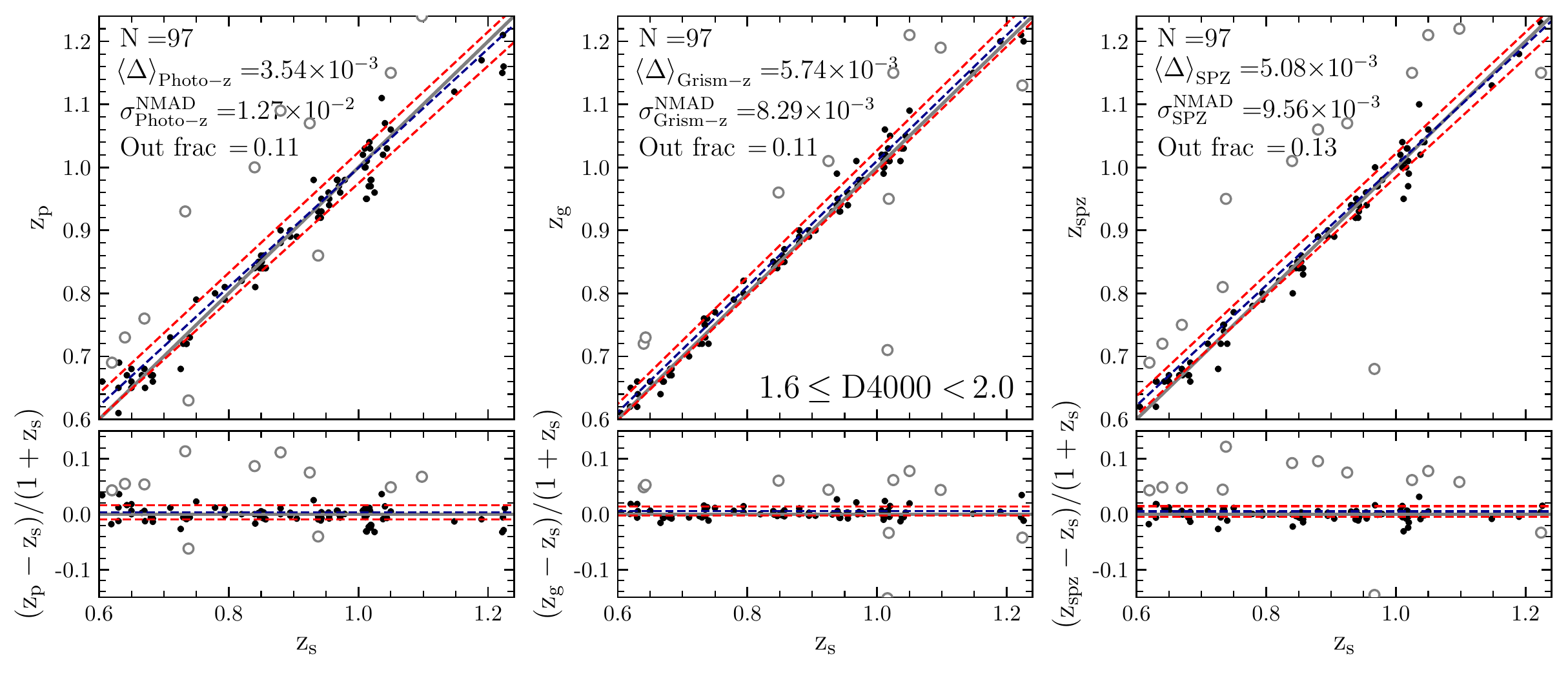}
\includegraphics[width=0.9\textwidth]{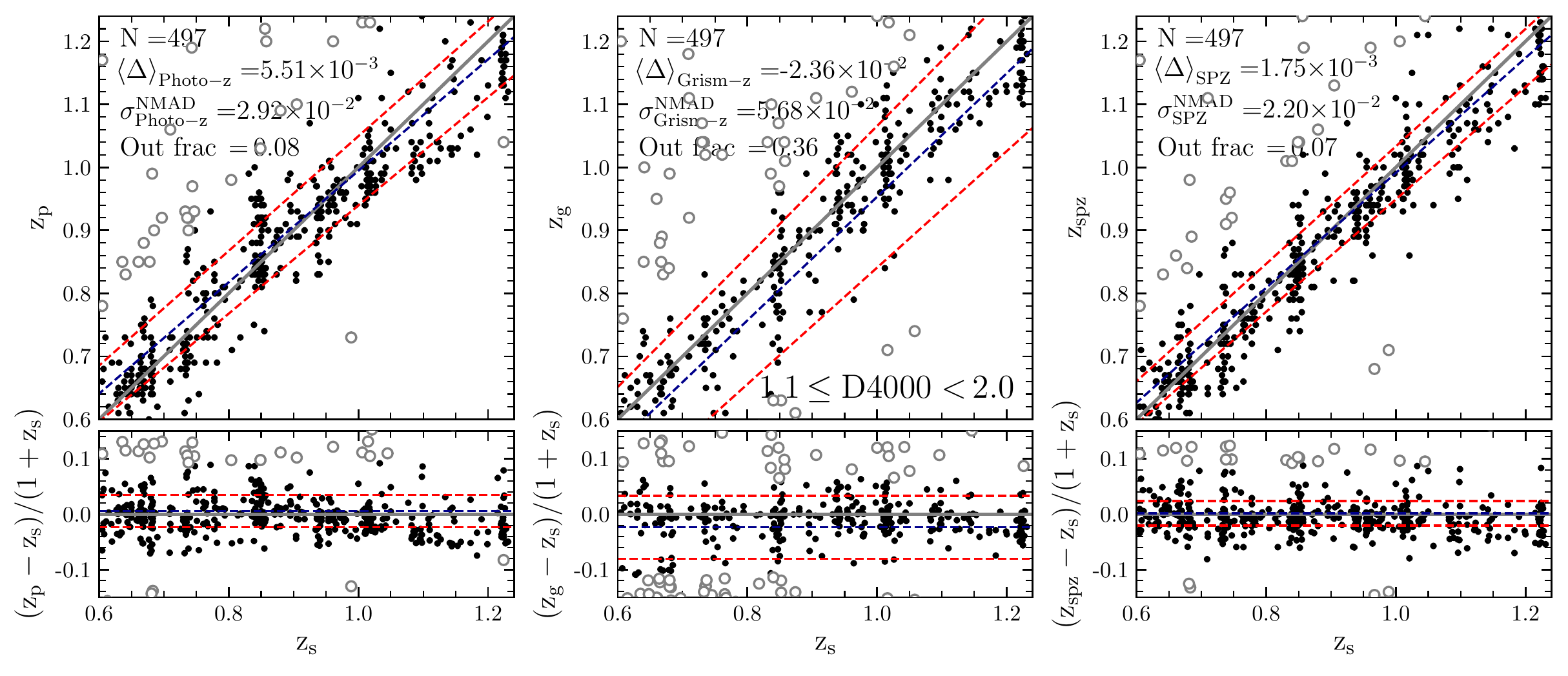}
\caption{Same as Figure \ref{fig:z_comparison_lowd4000} but for the two larger D4000 bins (top and middle) and the entire D4000 range (bottom).}
\label{fig:z_comparison_highd4000}
\end{figure*}

We derive redshift uncertainties analytically, by first deriving errors on the reduced $\chi^2$ \citep[see, e.g.,][]{Andrae2010, Hogg2010}, and then deriving the corresponding error on the redshift. The error on the reduced $\chi^2$ is given by $\sqrt{2/d.o.f.}$, where d.o.f. is the number of degrees of freedom. Our model has one free parameter, the vertical scaling factor $\alpha$, and the d.o.f. is given by the difference between number of observed flux points (N$\mathrm{_{data}}$) and the number of parameters in fit which gives d.o.f. = $\left(\mathrm{N_{data}} - 1\right)$. This also allows us to accurately estimate asymmetric uncertainties in cases where the $\chi^2$ minimum is asymmetric.

\begin{deluxetable*}{ c c c c c c c c c c c}
\tablecaption{Redshift statistics \label{tab:grismz_quants}}
\tablehead{
\colhead{D4000} & \colhead{N} & \multicolumn{3}{c}{Outlier fraction} & \multicolumn{2}{c}{Photo-z residuals} & \multicolumn{2}{c}{Grism-z residuals} & \multicolumn{2}{c}{SPZ residuals} \\
\colhead{range} & & \colhead{Photo-z} & \colhead{Grism-z} & \colhead{SPZ} & \colhead{Mean} & \colhead{$\mathrm{\sigma_{NMAD}}$} & \colhead{Mean} & \colhead{$\mathrm{\sigma_{NMAD}}$} & \colhead{Mean} & \colhead{$\mathrm{\sigma_{NMAD}}$}
}
\tablecolumns{9}
\startdata
1.1 $\leq$ D4000 $\leq$ 1.2 & 83   & 0.13 & 0.59  & 0.12 & +0.010     & 0.031 & $-$0.040    & 0.179 & $-$0.002  & 0.021 \\
1.2 $\leq$ D4000 $\leq$ 1.3 & 115 & 0.07 & 0.35  & 0.07 & $-$0.003  & 0.034 & $-$0.098 & 0.107 & $-$0.010  & 0.027 \\
1.3 $\leq$ D4000 $\leq$ 1.4 & 108 & 0.05 & 0.35  & 0.04 & +0.014     & 0.039 & $-$0.001  & 0.088  & +0.014     & 0.025 \\
1.4 $\leq$ D4000 $\leq$ 1.6 & 94   & 0.06 & 0.22  & 0.04 & +0.003     & 0.027 & +0.025     & 0.040 & +0.002   & 0.023 \\
1.6 $\leq$ D4000 $\leq$ 2.0 & 97   & 0.11 & 0.11   & 0.13 & +0.004     & 0.013 & +0.006     & 0.008 & +0.005     & 0.010 \\
Full D4000 range:  &  &  &  &  &  &  &  &  &  & \\
1.1 $\leq$ D4000 $\leq$ 2.0 & 497 & 0.08 & 0.36  & 0.07    & +0.006      & 0.029   & $-$0.024  & 0.057 & +0.002     & 0.022 \\
\enddata
\tablecomments{Quantifying the accuracy of the three different redshifts while stepping through D4000 bins.}
\end{deluxetable*}

Figure \ref{fig:example} shows four example spectra and photometry, with varying D4000 values, to show results from our SED fitting routine. It can be seen that our fitting process gives decent results for both 4000\AA\ (top row) and Balmer breaks (bottom row). Although it does not appear to affect the fitting results, we note that the $\chi^2$ values for SPZ and photo-z are much larger than the optimal value of 1.0. This is likely due to  systematic errors in the absolute photometric calibration, because our photometric data come from multiple different instruments and observatories. It can also be seen that the residuals for the grism data are scattered tightly around zero, whereas the photometric points can have much larger residuals. As another visual check, we also compute the p(z) curves for each galaxy. This is done by first converting the $\chi^2$ map to a likelihood map, $\mathcal{L}\,{\sim}\, e^{-\chi^2/2}$, which is then marginalized over all model parameters to convert it to a redshift probability distribution, i.e., a p(z) curve. In Table \ref{tab:catalog} we provide our redshift estimates and their uncertainties along with the other relevant parameters.

\section{The dependence of redshift accuracy on D4000}
\label{sec:accuracy}
We now investigate the dependence of the accuracy of the three types of redshifts on the 4000\AA\ break strength, D4000. Our derived redshifts are compared to ground-based spectroscopic redshifts which are often based on emission lines. 

Figures \ref{fig:z_comparison_lowd4000} and \ref{fig:z_comparison_highd4000} show this comparison of accuracy between the photometric (broad-band), grism (only), and spectrophotometric (broad-band plus grism) redshifts while stepping through bins of increasing D4000. We quantify the outlier fraction and the mean and spread of our sample for our three types of redshifts, for different bins of D4000, in Table \ref{tab:grismz_quants}. The mean of the residuals is given by $\left< \mathrm{ \Delta z/(1 + z_{s}) } \right>$, where $\Delta z$ is $\mathrm{(z_{p;g;spz} - z_{s})}$. The spread in the distribution of redshift residuals is measured by using the Normalized Median Absolute Deviation \citep[$\mathrm{\sigma_{NMAD}}$; see, for e.g.,][]{Brammer2008}. The $\mathrm{\sigma_{NMAD}}$ is given by:

\begin{equation}
\label{eq:sigma_nmad}
\mathrm{\sigma_{NMAD}} = 1.48 \times \mathrm{median\left(\frac{\Delta z - median(\Delta z)}{1+z_{s}}\right)}.
\end{equation}
A redshift value is defined as an outlier when its $\left| \mathrm{\Delta z / (1 + z_{s})} \right|$ value is greater than 3$\mathrm{\sigma_{NMAD}}$ away from the mean. For the sake of consistency we use the $\mathrm{\sigma^{phot}_{NMAD}}$, for photometric redshifts, as the $\sigma_\mathrm{NMAD}$ in the definition of outlier. The entire sample of \totalcatalog\ galaxies over the D4000 range of $1.1 \leq \mathrm{D}4000 < 2.0$  is shown in the bottom three panels in Figure \ref{fig:z_comparison_highd4000}. 

\begin{figure*}
\centering
\includegraphics[width=0.94\textwidth]{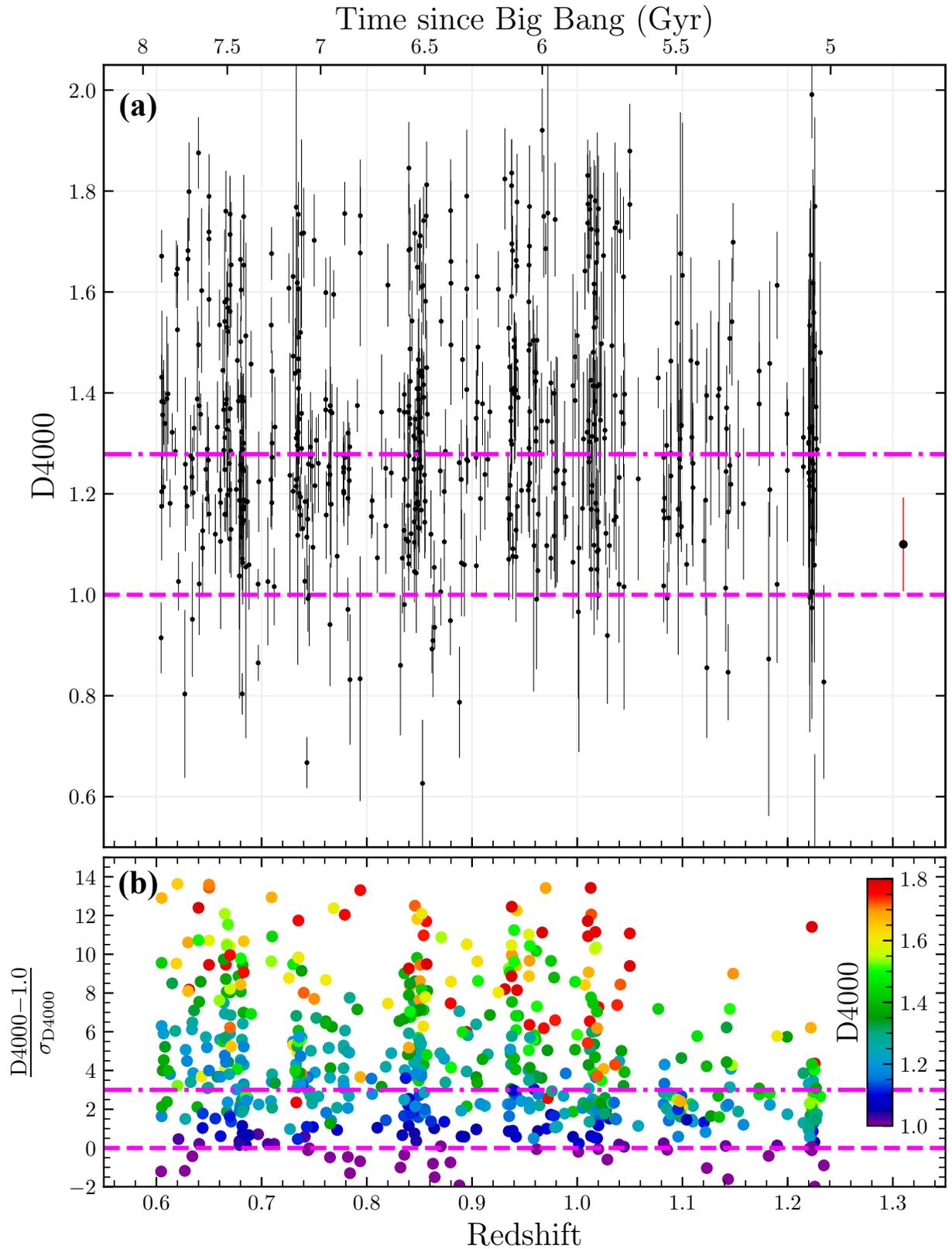}
\caption{(a) The distribution of $\mathrm{D}4000$ values with spectroscopic redshift. The black points show the $\mathrm{D}4000$ measurements from our PEARS data for all the 602 galaxies which were within our redshift range and passed the Net Spectral Significance and contamination cuts. The horizontal pink dashed line at $\mathrm{D}4000 = 1.0$ shows the $\mathrm{D}4000$ value for a flat spectrum, in f$_\nu$. The pink dash-dot line is 3$\sigma$ away from 1.0, where 1$\sigma$ is the average error-bar for all D4000$<$1.3 shown by the point with the red error-bar. (b) The significance of the D4000 measurement, as measured by $(D4000 - 1.0)/\sigma_{D4000}$, vs.\ redshift. The points are colored with their D4000 values as shown. The horizontal dashed and dash-dot lines show the null-measurement and 3$\sigma$ level, respectively.}
\label{fig:d4000_redshift}
\end{figure*}

It can be seen that the grism redshifts at values D4000$<$1.3 show a $\mathrm{\Delta z / (1 + z_{s})}$ offset at the level of $<-0.1$. This is due to a significant number of outliers toward the blue end of the redshift range. This occurs because, for galaxies with a relatively weaker 4000\AA/Balmer break -- and therefore lower significance (see following discussion and Figure \ref{fig:d4000_redshift}b) on the measured value of D4000 -- our fitting routine is driven to the blue edge of the redshift grid. This is also the reason why the middle subplot in the first two rows, of Figure \ref{fig:z_comparison_lowd4000}, has somewhat fewer points than the other two subplots. For D4000$\gtrsim$1.3 this offset is significantly diminished, and is similar for all three types of redshifts. From Figures \ref{fig:z_comparison_lowd4000} and \ref{fig:z_comparison_highd4000} and Table \ref{tab:grismz_quants}, it can be seen that the SPZ improves over the photo-z, as measured by the spread in the residuals, i.e., \ $\sigma_\mathrm{NMAD}$, for all D4000 values by $\sim$17--60\%. In the D4000 bin with the largest values, 1.6$\leq$D4000$<$2.0, the SPZ improves over the photo-z by 30\%.

A general trend for lower or similar SPZ outlier fractions as compared to photo-z outlier fractions can also be observed. For the bin with the largest D4000 values the outlier fraction for SPZs is $\sim$2\% higher than photo-z (i.e., the SPZ method has 2 additional objects which are counted as outliers). The photo-z method is expected to work well with the largest breaks and therefore shows a roughly equivalent outlier fraction for the largest D4000 bin. It therefore appears that the outlier fractions are less dependent on the method than the adopted definition of outlier.

\begin{figure}
\centering
\includegraphics[width=0.48\textwidth]{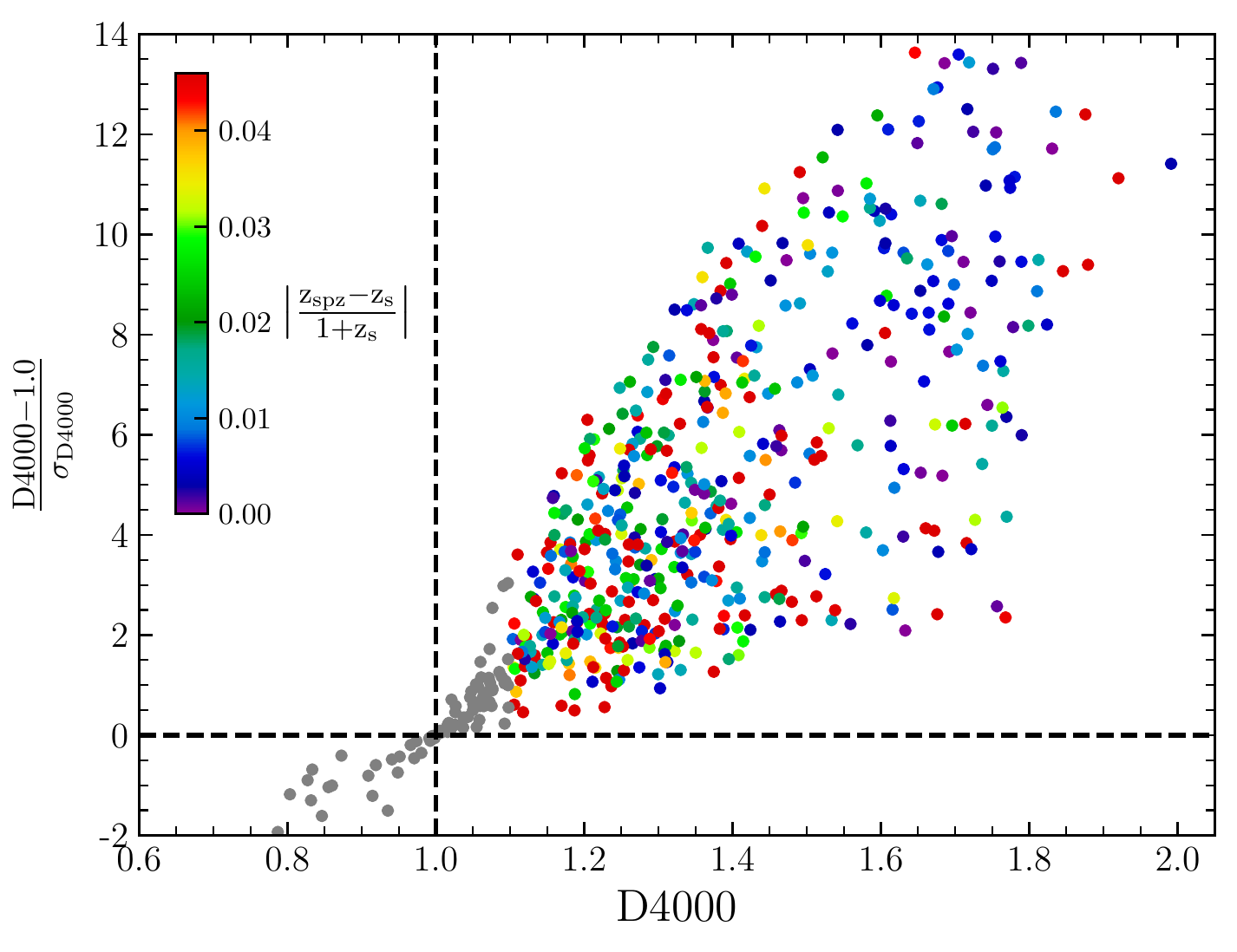}
\caption{The significance of the D4000 measurement vs.\ the D4000 value. The points are colored according with the accuracy of their spectrophotometric redshifts. The horizontal dashed line is the null-measurement level of D4000, and the vertical dashed line is D4000=1.0 i.e, a flat spectrum. The gray points are galaxy spectra with D4000$<$1.1 for which we did not estimate a spectrophotometric redshift with our routine. This shows that the signal-to-noise of D4000 can indicate the quality of redshift.}
\label{fig:d4000_sig}
\end{figure}

The dependence of redshift accuracy on D4000 also prompted us to attempt to quantify the value of D4000 at which the measurement of the break strength is significant enough for a redshift to be accurately estimated in Figure \ref{fig:d4000_redshift}b. For this, we quantify the D4000 measurement significance by $(\mathrm{D}4000 - 1.0)/\sigma_\mathrm{D4000}$, which is plotted vs.\ redshift in Figure \ref{fig:d4000_redshift}b, where the points are colored with their D4000 value ($\sigma_\mathrm{D4000}$ is the error in our D4000 measurement). In Figure \ref{fig:d4000_redshift}a, we show the average error for all our D4000 measurements as the point with red error-bars. This average error is $\sim$0.1. We define the significance threshold then as the level of 3$\sigma$ above the flat spectrum line of D4000=1.0, shown as the pink dash-dot line at D4000$\simeq$1.3. This agrees well with the decreased offset in the grism redshifts which occurs at D4000$\geq$1.3 (Figures \ref{fig:z_comparison_lowd4000} and \ref{fig:z_comparison_highd4000}).  Figure \ref{fig:d4000_sig} shows this dependence between D4000 measurement significance and the redshift accuracy more explicitly. It can be seen that at larger D4000 values and larger significance of the D4000 measurements, the redshift accuracy improves dramatically, while at lower D4000 values and lower D4000 measurement significance, there exists a larger scatter in the redshift accuracy. 

In Figure \ref{fig:d4000_sig_vs_imag} we investigate the effect of source brightness on the significance of the D4000 measurement. The figure shows a strong correlation between $i_{AB}$ mag and $({\rm D4000} - 1.0)/\sigma_{\rm D4000}$. It can be seen that at fluxes fainter than $i_{AB}\sim23\text{--}24$ mag, there are far more galaxies with lower D4000 significance, i.e., $({\rm D4000} - 1.0)/\sigma_{\rm D4000} < 3.0$ than with higher D4000 significance.

This idea of D4000 significance is analogous to the measurement of the significance of emission lines when determining redshifts based on emission lines. Essentially, for an accurate redshift to be measured, the emission line must be measured at a significance of \emph{at least} 3$\sigma$ above the continuum level. We conclude here that a similar requirement is needed on the D4000 measurement for accurate redshifts based on absorption features. Therefore, using the robustness of D4000 measurements will be a useful tool for future WFIRST and Euclid redshift surveys.

\section{Estimates of object number densities for future surveys}
\label{sec:num_density}
The Wide Field Infrared Survey Telescope (WFIRST) was ranked as the highest priority space mission in the Astro2010 decadal survey, ``New Worlds New Horizons'' \citep{Blandford2010}. One of the primary drivers of both WFIRST and the European Space Agency's Euclid mission is to measure the growth of structure and cosmic expansion over a large period of cosmic history. To achieve this, accurate redshift measurements are necessary -- within 0.1\% accuracy for baryon acoustic scale measurements (BAO) and ${\sim}\text{1-3}\%$ for weak lensing and galaxy overdensity measurements. In this section, we estimate the expected number density of objects by future redshift surveys that will achieve a redshift accuracy of $\sim$2\% or better based on the 4000\AA/Balmer breaks.

As a first step to estimate the number density for future redshift surveys, we investigate any possible evolution of the 4000\AA\ break strength, D4000, with redshift at intermediate redshifts. Figure \ref{fig:d4000_redshift}a shows the distribution of D4000 vs.\ redshift for all galaxies in our sample that had ground-based spectroscopic redshifts, and that passed our Net Spectral Significance and contamination cuts. Figure \ref{fig:d4000_redshift}a shows that the strength of the 4000\AA\ break remains roughly constant between \pearszrange. It can also be seen that most of the galaxies at these redshifts have a red slope at $\sim$4000\AA, since they lie above the value of 1.0 (red dashed horizontal line) that represents a flat spectrum (in f$_\nu$) between 3750\AA\ and 4250\AA\ (equal integrated flux density in the two bandpasses used in the D4000 definition).

The method presented in this work relies on the presence of a discernible 4000\AA\ break. Hence, the number density of galaxies with accurate 4000\AA\ break redshifts observable by WFIRST and Euclid will depend on the fraction of galaxies that contain a discernible 4000\AA\ break at the redshifts being probed. 
For WFIRST and Euclid \citep{Laureijs2011}, the wavelength coverage of the grism is 1.0 to 1.93 $\mu$m (for the WFIRST Wide Field Instrument, WFI)\footnote{\url{https://wfirst.ipac.caltech.edu/sims/Param_db.html}}, and 0.92 to 1.85 $\mu$m (for the Euclid Near Infrared Spectrometer and Photometer, NISP)\footnote{\url{https://www.euclid-ec.org/?page_id=2490}}, respectively. These grism wavelength coverages translate to redshift ranges that are sensitive to the 4000\AA\ break, i.e., $1.67 \leq z \leq 3.45$ and $1.45 \leq z \leq 3.35$ for WFIRST and Euclid, respectively. These ranges are derived from wavelength coverage for the grisms and the D4000 definition (see Eq. \ref{eq:d4000}).

\begin{figure}
\centering
\includegraphics[width=0.48\textwidth]{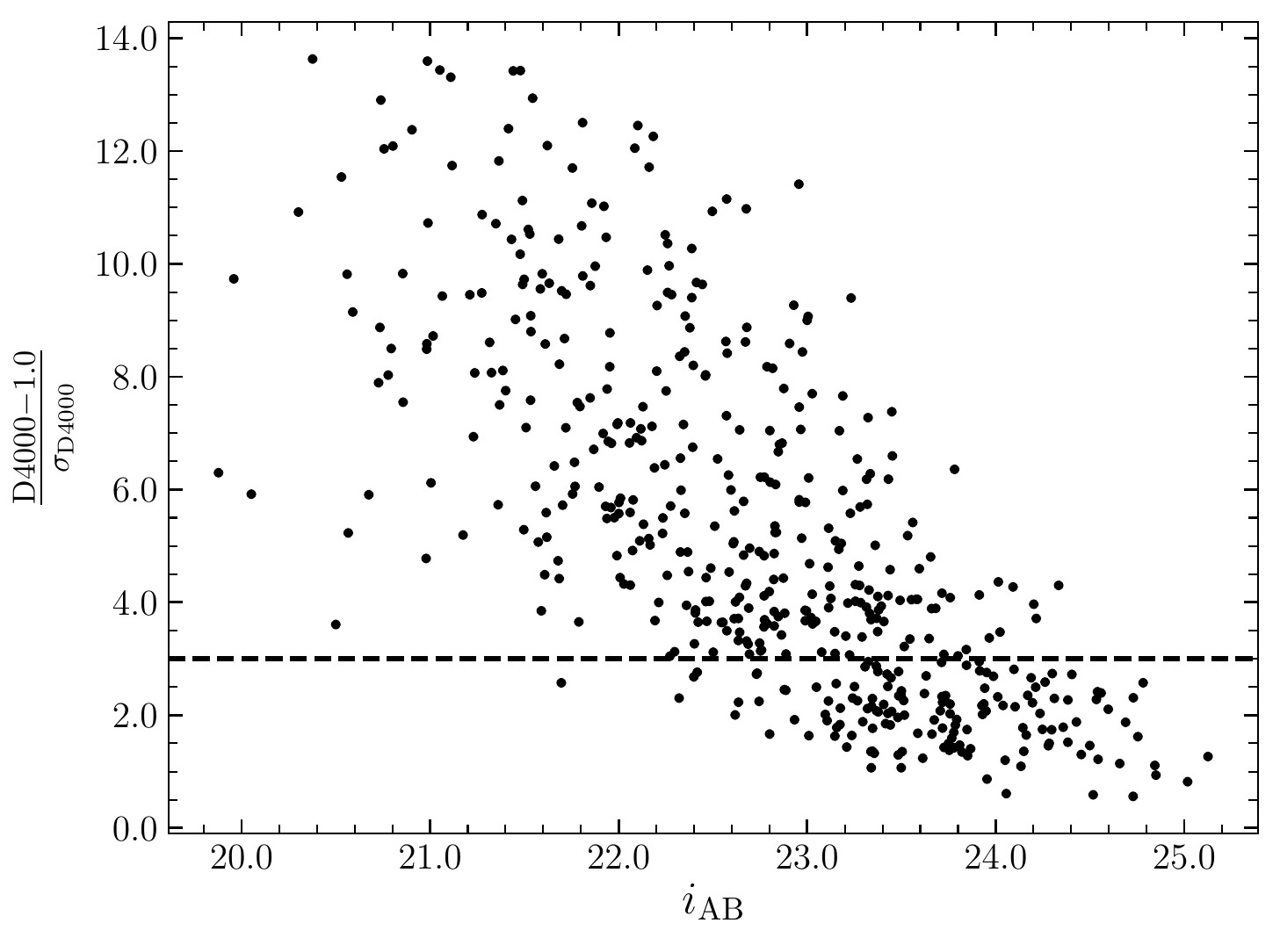}
\caption{Significance of the D4000 measurement vs.\ $i_{AB}$ magnitude. The horizontal dashed line is at ${\rm (D4000 - 1.0)}/{\sigma_{\rm D4000}}=3.0$. This plot includes all the \totalcatalog\ galaxies in our sample with $1.1 \leq \rm{D4000} < 2.0$.}
\label{fig:d4000_sig_vs_imag}
\end{figure}

Given the results from Figure \ref{fig:d4000_redshift}a --- the absence of a strong trend in D4000 with redshift for the intermediate redshift values we consider --- and also for simplicity, we assume that the fraction of galaxies with a 4000\AA\ or Balmer break available to provide accurate redshifts is roughly constant between the WFIRST and Euclid redshift ranges. We are allowing for the possibility that the abundance of 4000\AA\ break galaxies should drop at higher redshifts, but the abundance of Balmer break galaxies (due to post starburst populations dominated by A-type stars) should increase due to the increased overall star formation activity in the Universe \citep[e.g.,][]{Madau2014}.

Based on the results from Figure \ref{fig:d4000_sig_vs_imag}, for the remainder of this work, we will only consider the subsample of galaxies within our sample that are brighter than $i_{AB}$=24 mag. There are 464 galaxies, representing $\sim$72\% of all galaxies in the PEARS area (464 out of 646 galaxies), that are within \pearszrange\ and contain a discernible 4000\AA/Balmer break, i.e., D4000$>$1.1, and also have $i_{AB}$$\leq$24 mag. We assume that this is the fraction of galaxies within any given area that will contain 4000\AA/Balmer breaks available to determine accurate redshifts.

We estimate the number density of objects with accurate 4000\AA/Balmer break redshifts by integrating measured luminosity functions (LFs) over the WFIRST and Euclid redshift ranges covering the 4000\AA\ break. The measured LFs come from \citet{Kelvin2014}, who report LFs measured for elliptical galaxies (among others) at local redshifts, using data from the GAMA survey \citep{Driver2011}.
The LFs are assumed to follow a Schechter form \citep{Schechter1976} parameterized by the characteristic magnitude $\rm M^*$, the characteristic number density $\phi^*$ [$\rm Mpc^{-3} mag^{-1}$], and the faint-end slope $\alpha$. While the choice of LF does impact the predicted number counts, the effects of LF-evolution will be minor, and the largest uncertainty is currently the unknown survey completeness.

Following the prescription in \citet{Gardner1998}, we integrate the adopted LF and differential volume element, giving the number density out to the specified magnitude limit. The BC03 SED we used has the following stellar population parameters: age, $\mathrm{t = 4\, Gyr}$, exponential SFH  timescale, $\tau = 0.1\, \mathrm{Gyr}$, $\mathrm{A_V = 0}$, and solar metallicity. To predict the number density of galaxies that contain a discernible 4000\AA\ break and also allow for a redshift estimate with accuracy $\left| \mathrm{\Delta z/(1+z_{s})} \right| \leq 0.02$, we multiply the number densities obtained from the LF integration with the fraction of galaxies with D4000$>$1.1 (denoted $f_{D4000}$) and the fraction of galaxies with $\left| \mathrm{\Delta z/(1+z_{s})} \right| \leq 0.02$ (denoted $f_{acc}$) determined from our sample. These fractions are $f_{D4000}=0.72$ (see preceding discussion) and  $f^{SPZ}_{acc}=0.575$ (286 out of 497 galaxies) for spectrophotometric redshifts or $f^{zg}_{acc}=0.36$ (180 out of 497 galaxies) for grism-only redshifts.

Therefore, based on the calculation described above, we predict a total of $\sim$700--4400 galaxies/degree$^2$ which can be used to obtain redshifts accurate to $\left| \mathrm{\Delta z/(1+z_{s})} \right| \leq 0.02$, within the WFIRST and Euclid redshift ranges to a limiting depth of $i_{AB}$=24 mag. 
Our predicted number density of galaxies with accurate continuum derived redshifts is comparable to the expected number density of emission line redshifts. For example, our expected number density agrees reasonably well, at the higher end of our prediction, with the expected number density of H$\alpha$ emitting galaxies predicted by \citet{Merson2018}, who find a number density between 3900 and 4800 galaxies/degree$^2$ for a Euclid-like survey, and a number density between 10400 and 15200 galaxies/degree$^2$ for a WFIRST-like survey. Our expected number density also agrees well with estimates of line-emitting galaxies from the WFC3 Infrared Spectroscopic Parallels (WISP) survey collaboration \citep[see for e.g.,][]{Colbert2013, Mehta2015} 
and also with the number counts of line emitters at $0.9 < z < 1.8$ estimated by \citet{Valentino2017}.

\section{Conclusion}
\label{sec:conclusion}
We note three important aspects of our continuum derived redshifts relevant to future redshift surveys --\\ 
(i) {\bf Complementarity with emission line redshifts}: As we have shown, the expected number density of galaxies with redshifts derived from the 4000\AA/Balmer breaks is comparable to that of galaxies with H$\alpha$ based redshifts. Our redshift fitting method relies on absorption features in the continuum to minimize $\chi^2$, while also accounting for the effects of correlated data in the grism spectra and galaxy morphology. Since the two methods rely on very different features present in galaxy spectra, these methods individually access very different galaxy populations. Therefore, the two methods combined can comprehensively sample the galaxy population. We have also shown, that the D4000 measurement significance can be used as a proxy for expected continuum-based redshift accuracy.\\
(ii) {\bf Redshifts based on grism data alone}: For galaxies that contain strong emission lines, using grism data alone, a redshift accuracy of $\sim$0.1\% can be achieved, by employing 2-dimensional grism spectra to detect individual emission line regions in galaxies \citep{Pirzkal2018}. We have shown that continuum-based redshifts derived using only grism data, for galaxies \emph{without} strong emission lines, can still achieve an accuracy of $\sigma^{\rm NMAD}_{\rm \Delta z/(1+z)}$$\sim$6\% (Table \ref{tab:grismz_quants}), down to $i_{AB}$$\sim$23--24 mag. This is when using ACS/G800L spectra that have R$\simeq$100 and a typical rest-frame coverage around the 4000\AA\ break of $\sim$1500\AA. This is especially important, given that much of the area covered by WFIRST and Euclid will not have supporting 12-band (or more) photometry, and therefore must rely on grism-based redshifts alone.\\
(iii) {\bf Grism redshifts for fainter continua}: Accurate grism continuum redshifts can be achieved for {\it continua} that are fainter ($i_{AB}$=24 mag for this paper) than those that can be done from the ground. Given the steep slope in the luminosity function at the faint end \citep[e.g.,][]{Finkelstein2015}, this is particularly important because sampling fainter magnitudes will allow for obtaining continuum-based redshifts for much larger numbers of galaxies.

While programs targeting H$\alpha$ and [OIII]$\lambda$5007 lines are being planned for both the WFIRST \citep[see e.g., High Latitude Survey,][]{Spergel2015} and Euclid missions, we show that redshifts obtained with the 4000\AA/Balmer breaks can also be accurate to at least $\lesssim$2\%. We argue that -- (i) since the expected number densities of objects with redshifts based on the 4000\AA/Balmer breaks and objects with emission line redshifts are comparable, and (ii) since grism continuum redshifts can be done from space to fainter continuum levels compared to continuum-based redshifts from the ground (e.g., this work goes as faint as $i_{AB}$=24 mag) -- continuum-based redshifts can thus provide redshifts for galaxies which will not have emission line based redshifts from grism observations with WFIRST and Euclid, and therefore contribute additional redshifts which would otherwise not be available.

\acknowledgments
B.A.J.\ is grateful to Drs.\ Philip Appleton, Sanchayeeta Borthakur, Sangeeta Malhotra and James Rhoads for helpful discussions on work done in this paper. The authors also thank the anonymous referee for their helpful comments and suggestions. The authors acknowledge Research Computing at Arizona State University for providing high performance computing resources that have contributed to the research results reported within this paper.\footnote{\url{http://www.researchcomputing.asu.edu}} RAW acknowledges JWST Interdisciplinary Scientist grants NAG5-12460, NNX14AN10G, and 80NSSC18 K0200 from NASA GSFC. This research has made use of NASA's Astrophysics Data System. This work makes use of observations taken by the PEARS treasury program with the NASA/ESA HST. This work also makes use of observations taken by the 3D-HST Treasury Program and CANDELS Multi-Cycle Treasury Program with the NASA/ESA HST. Some of the data presented in this paper were obtained from the Mikulski Archive for Space Telescopes (MAST) at the Space Telescope Science Institute. STScI is operated by the Association of Universities for Research in Astronomy, Inc., under NASA contract NAS 5-26555. This research has made use of the Python programming language along with the Numpy, Scipy, and Matplotlib packages. This research has made use of Astropy, a community-developed core Python package for Astronomy \citep{astropy2018}.

\clearpage

\LongTables
\begin{deluxetable*}{l c r r c c c c r c c c}
\tablecaption{Redshift and D4000 catalog for our sample \label{tab:catalog}}
\tabletypesize{\scriptsize}
\tablehead{
\colhead{PearsID} & \colhead{Field} & \colhead{RA} & \colhead{DEC} & \colhead{z$_{\rm spec}$} & \colhead{z$_{\rm phot}$} & \colhead{z$_{\rm grism}$} & \colhead{z$_{\rm SPZ}$} & \colhead{$\mathcal{N}$} & \colhead{D4000} & \colhead{$\sigma_\mathrm{D4000}$} & \colhead{\it{i}$(\mathrm{AB})$}
}
\startdata
89573    & GOODS-N   & 189.2496182   & 62.274447   & 1.012   & 0.95$\substack{+0.11 \\ -0.00}$   & 1.06$\substack{+0.00 \\ -0.02}$   & 0.95$\substack{+0.18 \\ -0.00}$   & 136.31   & 1.7639   & 0.12   & 23.27 \\
96475    & GOODS-N   & 189.3574708   & 62.287353   & 0.915   & 0.92$\substack{+0.00 \\ -0.00}$   & 0.77$\substack{+0.00 \\ -0.00}$   & 0.92$\substack{+0.00 \\ -0.09}$   & 216.75   & 1.2684   & 0.06   & 22.66 \\
66753    & GOODS-N   & 189.2062393   & 62.235220   & 0.752   & 0.83$\substack{+0.06 \\ -0.01}$   & 0.74$\substack{+0.00 \\ -0.27}$   & 0.83$\substack{+0.03 \\ -0.03}$   & 395.31   & 1.3061   & 0.05   & 21.87 \\
94085    & GOODS-N   & 189.2999221   & 62.283150   & 1.142   & 1.03$\substack{+0.01 \\ -0.02}$   & 1.12$\substack{+0.00 \\ -0.01}$   & 1.04$\substack{+0.00 \\ -0.02}$   & 360.69   & 1.3292   & 0.05   & 22.77 \\
33414    & GOODS-N   & 189.1540648   & 62.175396   & 1.016   & 0.97$\substack{+0.02 \\ -0.01}$   & 0.71$\substack{+0.65 \\ -0.01}$   & 1.00$\substack{+0.00 \\ -0.03}$   & 89.50    & 1.6158   & 0.25   & 23.43 \\
74987    & GOODS-N   & 189.2407802   & 62.248600   & 0.849   & 0.92$\substack{+0.00 \\ -0.00}$   & 0.97$\substack{+0.01 \\ -0.02}$   & 0.92$\substack{+0.00 \\ -0.03}$   & 293.54   & 1.3627   & 0.05   & 22.12 \\
83499    & GOODS-N   & 189.1973120   & 62.274485   & 0.871   & 0.87$\substack{+0.00 \\ -0.00}$   & 0.87$\substack{+0.03 \\ -0.01}$   & 0.87$\substack{+0.00 \\ -0.00}$   & 603.06   & 1.5416   & 0.05   & 21.28 \\
125098   & GOODS-N   & 189.3846421   & 62.342237   & 1.223   & 1.21$\substack{+0.00 \\ -0.00}$   & 1.30$\substack{+0.00 \\ -0.00}$   & 1.23$\substack{+0.00 \\ -0.00}$   & 319.61   & 1.9912   & 0.09   & 22.96 \\
124386   & GOODS-N   & 189.3902768   & 62.344510   & 0.841   & 0.86$\substack{+0.01 \\ -0.02}$   & 0.93$\substack{+0.02 \\ -0.09}$   & 0.86$\substack{+0.01 \\ -0.06}$   & 222.50   & 1.4873   & 0.07   & 22.80 \\
86522    & GOODS-N   & 189.3232826   & 62.268957   & 0.710   & 0.68$\substack{+0.02 \\ -0.00}$   & 1.32$\substack{+0.02 \\ -0.04}$   & 0.64$\substack{+0.04 \\ -0.01}$   & 269.71   & 1.1821   & 0.05   & 22.86 \\
76405    & GOODS-N   & 189.1780463   & 62.250802   & 0.697   & 0.92$\substack{+0.01 \\ -0.01}$   & 0.44$\substack{+0.01 \\ -0.01}$   & 0.83$\substack{+0.00 \\ -0.01}$   & 119.74   & 1.2246   & 0.10   & 23.11 \\
92378    & GOODS-N   & 189.1805813   & 62.281280   & 0.944   & 1.00$\substack{+0.05 \\ -0.03}$   & 1.04$\substack{+0.01 \\ -0.00}$   & 1.02$\substack{+0.09 \\ -0.04}$   & 288.62   & 1.3911   & 0.06   & 22.06 \\
82816    & GOODS-N   & 189.3380698   & 62.262313   & 0.778   & 0.82$\substack{+0.01 \\ -0.00}$   & 0.70$\substack{+0.01 \\ -0.00}$   & 0.78$\substack{+0.01 \\ -0.00}$   & 154.46   & 1.2012   & 0.07   & 22.69 \\
58856    & GOODS-N   & 189.1602768   & 62.220296   & 0.635   & 0.58$\substack{+0.00 \\ -0.01}$   & 0.40$\substack{+0.00 \\ -0.00}$   & 0.60$\substack{+0.01 \\ -0.00}$   & 287.12   & 1.2024   & 0.05   & 22.41 \\
70857    & GOODS-N   & 189.1161525   & 62.246957   & 0.680   & 0.70$\substack{+0.01 \\ -0.02}$   & 0.99$\substack{+0.01 \\ -0.03}$   & 0.71$\substack{+0.00 \\ -0.00}$   & 547.56   & 1.3923   & 0.05   & 21.33 \\
112509   & GOODS-N   & 189.4184551   & 62.314892   & 0.955   & 1.01$\substack{+0.01 \\ -0.06}$   & 1.46$\substack{+0.01 \\ -0.01}$   & 1.00$\substack{+0.03 \\ -0.05}$   & 141.09   & 1.3632   & 0.09   & 23.03 \\
79185    & GOODS-N   & 189.1656552   & 62.263279   & 0.848   & 0.84$\substack{+0.00 \\ -0.00}$   & 0.96$\substack{+0.00 \\ -0.11}$   & 0.85$\substack{+0.00 \\ -0.00}$   & 528.87   & 1.6492   & 0.05   & 21.36 \\
54598    & GOODS-N   & 189.2245564   & 62.215020   & 0.642   & 0.64$\substack{+0.03 \\ -0.02}$   & 0.61$\substack{+0.01 \\ -0.01}$   & 0.64$\substack{+0.01 \\ -0.03}$   & 694.02   & 1.3576   & 0.04   & 20.98 \\
121302   & GOODS-N   & 189.3372338   & 62.332220   & 0.778   & 0.76$\substack{+0.00 \\ -0.02}$   & 0.30$\substack{+0.00 \\ -0.00}$   & 0.72$\substack{+0.01 \\ -0.00}$   & 311.43   & 1.2509   & 0.05   & 22.16 \\
119879   & GOODS-N   & 189.3599907   & 62.329089   & 1.010   & 1.00$\substack{+0.00 \\ -0.00}$   & 1.00$\substack{+0.00 \\ -0.00}$   & 1.00$\substack{+0.00 \\ -0.00}$   & 284.78   & 1.7751   & 0.07   & 22.50 \\
41630    & GOODS-N   & 189.1987192   & 62.188955   & 1.223   & 1.20$\substack{+0.12 \\ -0.06}$   & 1.12$\substack{+0.00 \\ -0.00}$   & 1.23$\substack{+0.09 \\ -0.09}$   & 111.73   & 1.2653   & 0.14   & 23.17 \\
98997    & GOODS-N   & 189.3637473   & 62.293080   & 1.011   & 0.95$\substack{+0.02 \\ -0.02}$   & 0.95$\substack{+0.06 \\ -0.00}$   & 0.95$\substack{+0.01 \\ -0.00}$   & 143.34   & 1.3447   & 0.08   & 23.12 \\
86203    & GOODS-N   & 189.1572727   & 62.268766   & 0.841   & 0.81$\substack{+0.01 \\ -0.02}$   & 0.84$\substack{+0.00 \\ -0.00}$   & 0.80$\substack{+0.02 \\ -0.00}$   & 271.28   & 1.6855   & 0.08   & 22.32 \\
105455   & GOODS-N   & 189.2603162   & 62.305642   & 0.767   & 0.77$\substack{+0.01 \\ -0.01}$   & 0.50$\substack{+0.25 \\ -0.14}$   & 0.75$\substack{+0.00 \\ -0.01}$   & 258.67   & 1.3604   & 0.06   & 22.58 \\
66729    & GOODS-S   & 53.1363320   & $-$27.816537   & 0.671   & 0.65$\substack{+0.00 \\ -0.00}$   & 0.61$\substack{+0.00 \\ -0.00}$   & 0.65$\substack{+0.00 \\ -0.00}$   & 506.23   & 1.2847   & 0.04   & 21.95 \\
63902    & GOODS-S   & 53.1617459   & $-$27.824573   & 0.820   & 0.82$\substack{+0.00 \\ -0.00}$   & 0.82$\substack{+0.00 \\ -0.00}$   & 0.82$\substack{+0.00 \\ -0.00}$   & 221.30   & 1.6136   & 0.08   & 22.96 \\
81011    & GOODS-S   & 53.1603539   & $-$27.784002   & 0.954   & 0.90$\substack{+0.00 \\ -0.04}$   & 0.97$\substack{+0.00 \\ -0.01}$   & 0.90$\substack{+0.01 \\ -0.00}$   & 418.65   & 1.5806   & 0.05   & 21.92 \\
125952   & GOODS-S   & 53.0335122   & $-$27.701204   & 1.042   & 1.00$\substack{+0.01 \\ -0.02}$   & 1.29$\substack{+0.02 \\ -0.02}$   & 1.02$\substack{+0.02 \\ -0.01}$   & 227.76   & 1.3618   & 0.07   & 23.36 \\
115781   & GOODS-S   & 53.1484734   & $-$27.719482   & 1.222   & 1.15$\substack{+0.00 \\ -0.00}$   & 1.21$\substack{+0.00 \\ -0.00}$   & 1.15$\substack{+0.00 \\ -0.00}$   & 156.30   & 1.6731   & 0.11   & 23.01 \\
76592    & GOODS-S   & 53.1736667   & $-$27.797363   & 0.668   & 0.58$\substack{+0.07 \\ -0.02}$   & 0.36$\substack{+0.00 \\ -0.00}$   & 0.64$\substack{+0.00 \\ -0.00}$   & 132.67   & 1.1977   & 0.08   & 23.50 \\
26158    & GOODS-S   & 53.1785644   & $-$27.890211   & 0.650   & 0.66$\substack{+0.00 \\ -0.00}$   & 0.66$\substack{+0.00 \\ -0.00}$   & 0.66$\substack{+0.00 \\ -0.00}$   & 842.00   & 1.7049   & 0.05   & 20.99 \\
111741   & GOODS-S   & 53.0454611   & $-$27.728629   & 0.998   & 1.00$\substack{+0.03 \\ -0.02}$   & 1.10$\substack{+0.01 \\ -0.00}$   & 1.02$\substack{+0.01 \\ -0.03}$   & 330.77   & 1.4715   & 0.05   & 21.61 \\
31899    & GOODS-S   & 53.1398202   & $-$27.880603   & 1.077   & 1.12$\substack{+0.00 \\ -0.04}$   & 1.17$\substack{+0.00 \\ -0.11}$   & 1.11$\substack{+0.02 \\ -0.03}$   & 354.71   & 1.4295   & 0.06   & 22.06 \\
109151   & GOODS-S   & 53.1409173   & $-$27.736119   & 0.667   & 0.64$\substack{+0.00 \\ -0.00}$   & 0.65$\substack{+0.14 \\ -0.00}$   & 0.64$\substack{+0.00 \\ -0.00}$   & 520.90   & 1.5853   & 0.06   & 21.53 \\
132474   & GOODS-S   & 53.0865015   & $-$27.687486   & 0.683   & 0.64$\substack{+0.02 \\ -0.01}$   & 0.62$\substack{+0.00 \\ -0.00}$   & 0.66$\substack{+0.00 \\ -0.01}$   & 174.31   & 1.1884   & 0.07   & 22.73 \\
21592    & GOODS-S   & 53.1459236   & $-$27.901462   & 0.670   & 0.66$\substack{+0.00 \\ -0.00}$   & 0.83$\substack{+0.04 \\ -0.03}$   & 0.66$\substack{+0.00 \\ -0.01}$   & 443.13   & 1.5615   & 0.07   & 21.69 \\
34821    & GOODS-S   & 53.1652086   & $-$27.873947   & 1.096   & 1.04$\substack{+0.00 \\ -0.01}$   & 1.04$\substack{+0.00 \\ -0.01}$   & 1.03$\substack{+0.00 \\ -0.00}$   & 249.11   & 1.1185   & 0.06   & 22.62 \\
109794   & GOODS-S   & 53.1431093   & $-$27.730585   & 0.667   & 0.65$\substack{+0.00 \\ -0.00}$   & 1.49$\substack{+0.00 \\ -1.11}$   & 0.64$\substack{+0.00 \\ -0.00}$   & 519.81   & 1.3874   & 0.05   & 21.24 \\
118100   & GOODS-S   & 53.0703038   & $-$27.717853   & 0.644   & 0.62$\substack{+0.02 \\ -0.01}$   & 0.67$\substack{+0.00 \\ -0.00}$   & 0.67$\substack{+0.00 \\ -0.00}$   & 159.11   & 1.1269   & 0.07   & 23.16 \\
135087   & GOODS-S   & 53.0807722   & $-$27.681003   & 0.682   & 0.99$\substack{+0.00 \\ -0.00}$   & 0.46$\substack{+0.23 \\ -0.01}$   & 0.98$\substack{+0.00 \\ -0.00}$   & 714.63   & 1.3917   & 0.04   & 21.07 \\
19226    & GOODS-S   & 53.1412561   & $-$27.907070   & 0.663   & 0.65$\substack{+0.01 \\ -0.01}$   & 1.32$\substack{+0.00 \\ -0.00}$   & 0.73$\substack{+0.00 \\ -0.00}$   & 324.85   & 1.4449   & 0.08   & 22.23 \\
118459   & GOODS-S   & 53.0952520   & $-$27.717270   & 0.671   & 0.68$\substack{+0.00 \\ -0.01}$   & 0.38$\substack{+0.05 \\ -0.01}$   & 0.65$\substack{+0.01 \\ -0.00}$   & 104.73   & 1.1295   & 0.10   & 23.35 \\
108892   & GOODS-S   & 53.0779324   & $-$27.736875   & 0.958   & 1.00$\substack{+0.00 \\ -0.05}$   & 0.90$\substack{+0.05 \\ -0.01}$   & 0.94$\substack{+0.03 \\ -0.04}$   & 144.69   & 1.5038   & 0.09   & 22.61 \\
47038    & GOODS-S   & 53.1338939   & $-$27.851542   & 0.683   & 0.72$\substack{+0.01 \\ -0.04}$   & 0.31$\substack{+0.01 \\ -0.00}$   & 0.72$\substack{+0.01 \\ -0.06}$   & 153.20   & 1.3004   & 0.10   & 22.30 \\
89031    & GOODS-S   & 53.1016142   & $-$27.773408   & 0.895   & 0.90$\substack{+0.00 \\ -0.00}$   & 0.89$\substack{+0.01 \\ -0.00}$   & 0.90$\substack{+0.00 \\ -0.00}$   & 150.49   & 1.7899   & 0.13   & 22.60 \\
96844    & GOODS-S   & 53.1523301   & $-$27.761868   & 0.710   & 0.41$\substack{+0.00 \\ -0.01}$   & 0.31$\substack{+0.01 \\ -0.00}$   & 0.31$\substack{+0.01 \\ -0.00}$   & 165.48   & 1.2724   & 0.07   & 23.33 \\
113678   & GOODS-S   & 53.0611747   & $-$27.726963   & 0.910   & 0.97$\substack{+0.01 \\ -0.01}$   & 1.49$\substack{+0.00 \\ -1.13}$   & 0.94$\substack{+0.01 \\ -0.00}$   & 142.71   & 1.3786   & 0.08   & 23.44 \\
90546    & GOODS-S   & 53.1549743   & $-$27.768908   & 1.050   & 1.06$\substack{+0.00 \\ -0.00}$   & 1.09$\substack{+0.01 \\ -0.00}$   & 1.06$\substack{+0.00 \\ -0.00}$   & 577.57   & 1.7734   & 0.07   & 21.86 \\
83248    & GOODS-S   & 53.1616316   & $-$27.780247   & 0.619   & 0.59$\substack{+0.01 \\ -0.00}$   & 0.62$\substack{+0.00 \\ -0.00}$   & 0.59$\substack{+0.00 \\ -0.00}$   & 588.09   & 1.6353   & 0.07   & 21.70 \\
\enddata
\tablecomments{Redshift estimates from our galaxy sample. We also provide the Net Spectral Significance ($\mathcal{N}$), D4000, and the {\it i}-band magnitude for each galaxy. Only part of the table is shown here to illustrate its content; the entire machine readable table for \totalcatalog\ galaxies is available from the journal website.}
\end{deluxetable*}
\clearpage

\appendix
\section{A.\ Errors on the D4000 measurement}
\label{app:d4000_err}
\noindent We use a simple trapezoidal rule to estimate the definite integrals to measure the continuum flux within the two bandpasses employed in the measure of D4000. The error on the D4000 measurement is obtained analytically using equation \ref{eq:d4000_err}, where f$_-$ and f$_+$ are the measurements of integrated flux in the bluer and red bandpasses in the definition of D4000, respectively. The flux measurements, f$_-$ and f$_+$, are given by equations \ref{eq:flux_low} and \ref{eq:flux_high}, respectively, and $\sigma_-$ and $\sigma_+$ are their respective errors which are given by equations \ref{eq:err_low} and \ref{eq:err_high}. In the equations below, $\mathrm{f_{\lambda}}$ and $\mathrm{\sigma_{\lambda}}$ are the values of the flux and its corresponding error respectively at the stated $\lambda$:

\begin{equation}
\mathrm{D4000} = \frac{f_+}{f_-}
\end{equation}

\begin{equation}
\label{eq:d4000_err}
\sigma_{\mathrm{D}4000} = \frac{1}{f^2_-} \sqrt{\sigma^2_+ \, f^2_-\ +\ \sigma^2_- \, f^2_+}
\end{equation}

\begin{equation}
\label{eq:flux_low}
\mathrm{f}_- = \frac{(3950 - 3750)}{2\mathrm{N_-}} \left(\mathrm{f}_{3750} + \mathrm{f}_{3950} + 2\,\mathrm{f}_{\Sigma_-} \right)
\end{equation}

\begin{equation}
\label{eq:flux_high}
\mathrm{f}_+ = \frac{(4250 - 4050)}{2\mathrm{N_+}} \left(\mathrm{f}_{4050} + \mathrm{f}_{4250} + 2\,\mathrm{f}_{\Sigma_+} \right)
\end{equation}

\begin{equation}
\label{eq:err_low}
\sigma_- = \sqrt{\frac{(3950 - 3750)^2}{(2\mathrm{N_-})^2} \left(\sigma^2_{3750} + \sigma^2_{3950} + 4\,\sigma^2_{\Sigma_-} \right)}
\end{equation}

\begin{equation}
\label{eq:err_high}
\sigma_+ = \sqrt{\frac{(4250 - 4050)^2}{(2\mathrm{N_+})^2} \left(\sigma^2_{4050} + \sigma^2_{4250} + 4\,\sigma^2_{\Sigma_+} \right)}
\end{equation}

\noindent In the above equations, there are N$_-$+1 and N$_+$+1 spectral points in the blue and red wavelength intervals, respectively, over which the continuum flux is measured. In most spectra, where a flux measurement at the exact wavelengths of the bandpass limits did not exist, we used a simple linear interpolation to compute the flux measurement at the exact wavelength, using the flux measurements on either side of it. These flux points are referenced by their exact wavelengths in the above equations, i.e. $\mathrm{f}_{3750}$, $\mathrm{f}_{3950}$, $\mathrm{f}_{4050}$, $\mathrm{f}_{4250}$. Also, $\mathrm{f}_{\Sigma_-}$ and $\mathrm{f}_{\Sigma_+}$ are flux measurements excluding the end points in the blue and red wavelength intervals, respectively. 

\section{B.\ Net spectral significance}
\label{app:netsig}
The Net Spectral Significance ($\mathcal{N}$) is a way of measuring the amount of useful information in a spectrum \citep{Pirzkal2004}. It is defined as the maximum cumulative signal to noise ratio (SNR) in a spectrum.

The Net Spectral Significance is measured in the following way. The SNR at each flux point is measured and then sorted in descending order. This sorted array is used to create signal and noise arrays, with the first element in these arrays corresponding to the signal and noise from the highest SNR point in the spectrum. The second element in the signal array would then be the sum of the signals from the point with the highest SNR and the point with the second-highest SNR. The second element in the noise array are the noise values summed in quadrature for the same two points. The rest of the arrays are filled similarly.  

The signal and noise arrays are divided element-wise to make the final cumulative SNR array (shown in Eq. \ref{eq:cum_snr} as a sequence). The Net Spectral Significance is then the maximum value in this cumulative SNR array.

\begin{equation}
\label{eq:cum_snr}
\mathrm{Cumulative\ SNR}: \frac{S_1}{\sqrt{N^2_1}}, \frac{S_1 + S_2}{\sqrt{N^2_1 + N^2_2}}, \frac{S_1 + S_2 + S_3}{\sqrt{N^2_1 + N^2_2 + N^2_3}}, \mathrm{etc}\, \ldots
\end{equation}

\begin{equation}
\label{eq:netsig}
\mathcal{N} = \mathrm{max\, (Cumulative\ SNR)}
\end{equation}

\section{C.\ Covariance matrix estimation}
\label{app:covmat}

The covariance matrix is estimated using a squared exponential kernel \citep[see for example the brief pedagogical introduction given by][appendix A1]{Gibson2012}. We estimate individual elements of the inverse of the covariance matrix directly. The `ij'th element of the inverse of the covariance matrix `C$^{-1}$' is given by, 

\begin{equation}
\label{eq:covmat_ij}
\rm C^{-1}_{ij} =\delta_{ij}\frac{1}{\sigma^2_{ij}} +  (1 - \delta_{ij})\,\frac{1}{\theta_0}\,exp\left( - \frac{(\lambda_i -  \lambda_j)^2}{2\,\mathcal{L}^2} \right) .
\end{equation}

In the above equation, $\theta_0$ is the maximum covariance which is used as a normalization constant. We simply used the square of the maximum of errors on all flux points within a given spectrum. The effective correlation length is given by $\mathcal{L}$, and depends on the galaxy size along the dispersion direction. The wavelength corresponding to the `i'th flux point and the variance on the data is given by $\lambda_i$ and $\sigma^2_{ii}$, respectively. Here $\delta_{ij}$ is the Kronecker delta function which populates elements only on the diagonal of the matrix. In the ideal case of each grism data point being completely uncorrelated to any other data point in the grism spectrum, the inverse of the covariance matrix is a diagonal matrix containing only the reciprocals of the variances on the individual data points.

To arrive at the effective correlation length, $\mathcal{L}$, we fit the measured LSF of each galaxy with a Gaussian and set  $\mathcal{L} = 3\, \sigma_{LSF}$, where $\sigma_{LSF}$ is the best-fit Gaussian standard deviation for the LSF. This takes into account the correlation induced by the morphology of each galaxy along the dispersion direction. The above measure of the effective correlation length essentially indicates that  the data becomes uncorrelated $\pm3\,\sigma_{LSF}$ away from any individual data point in the grism spectrum.

Below we provide a short derivation for the vertical scaling factor given in equation \ref{eq:alpha}. The $\chi^2$ statistic is defined by $\chi^2 = \left( F - \alpha M \right)^T \, C^{-1} \, \left( F - \alpha M \right)$ which can be written as,

\begin{equation}
\chi^2 = \sum^N_{ij} \left( F_i - \alpha M_i \right) \left( F_j - \alpha M_j \right) \left( \frac{1}{\sigma^2_{ij}} \right).
\end{equation}

In the above equation, the `i'th flux and model elements are denoted by $F_i$ and $M_i$, respectively. The `ij'th element of $C^{-1}$ is denoted by $1/\sigma^2_{ij}$ which is the result of evaluating equation \ref{eq:covmat_ij}. For example, $C^{-1}_{11} = 1/\sigma^2_{11}$  where $\sigma^2_{11}$ is the variance on the `i'th flux point and $C^{-1}_{12} = 1/\sigma^2_{12} = (1/\theta_0)\, {\rm exp(-(\lambda_1 - \lambda_2)^2 / (2\,\mathcal{L}))}$. The size of the flux and model vectors is `N' elements and the size of $C^{-1}$ is N $\times$ N. We can now evaluate,

\begin{equation}
\frac{\partial \chi^2}{\partial \alpha} = \sum^N_{ij}\,  \left( \frac{1}{\sigma^2_{ij}} \right)\, \left[  \left( F_i - \alpha M_i \right).-M_j  +  \left( F_j - \alpha M_j \right).-M_i  \right] = 0,
\end{equation}

for the $\alpha$ that minimizes $\chi^2$. This then gives us,

\begin{equation}
\sum^N_{ij}\,  \left( \frac{1}{\sigma^2_{ij}} \right)\, \left[  -F_i M_j  -F_j M_i  + 2\alpha M_i M_j \right] = 0.
\end{equation}

Which directly leads to the $\alpha$ given by equation \ref{eq:alpha}.

\end{document}